%% file: SMP-19-007_temp.tex
\documentclass[11pt,twoside,a4paper,cmspaper,final,collab]{cms-tdr}
\def\svnVersion{c1daff6-D}\def\svnDate{2025/08/12}\def\cmsCernNoTag{CERN-EP-2025-134}\def\cmsCernDate{\today}\def\cmsMessage{Submitted to Physical Review D}
\begin{document}\cmsNoteHeader{SMP-19-007}

\newlength\cmsTabSkip\setlength{\cmsTabSkip}{1ex}
\pdfsuppresswarningpagegroup=1
\newcommand{\refeq}[1]{(\ref{#1})}
\newcommand{\PU}{{\HepParticle{U}{}{}}\Xspace}
\newcommand{\PKzLS}{{\HepParticle{\PK}{L,S}{0}}\Xspace} 
\newcommand{\muR}{\ensuremath{\mu_{\mathrm{R}}}\xspace}
\newcommand{\muF}{\ensuremath{\mu_{\mathrm{F}}}\xspace}
\newcommand{\CP} {\ensuremath{C\!P}\xspace}
\newcommand{\Br}{\mathcal{B}}
\newcommand{\fnr}{f_\text{nr}}
\newcommand{\axe}{(A \, \epsilon)}
\newcommand{\Nobs}{N^\text{obs}}
\newcommand{\Nbkg}{N^\text{bkg}}
\newcommand{\Nsig}{N^\text{sig}}
\newcommand{\Npr}{N^\text{pr}}
\newcommand{\Nnpr}{N^\text{npr}}
\newcommand{\looseCR}{_\text{loose CR}}
\newcommand{\loose}{_\text{loose}}
\newcommand{\PP}{\Pp{}\Pp}
\renewcommand{\LL}{\Pellp\Pellm}
\newcommand{\ZtoLL}{\PZ \to \LL}
\newcommand{\ZtoEE}{\PZ \to \EE}
\newcommand{\ZtoMM}{\PZ \to \MM}
\newcommand{\fL}{4\Pell}
\newcommand{\fM}{4\Pgm}
\newcommand{\tMtE}{2\Pgm 2\Pe}
\newcommand{\fE}{4\Pe}
\newcommand{\ZtofL}{\PZ \to \fL}
\newcommand{\ZtofM}{\PZ \to \fM}
\newcommand{\ZtotMtE}{\PZ \to \tMtE}
\newcommand{\ZtofE}{\PZ \to \fE}
\newcommand{\WW}{\PWp\PWm}
\newcommand{\WZ}{\PW\PZ}
\newcommand{\ZZ}{\PZ\PZ}
\newcommand{\ZZtofL}{\PZ\ZtofL}
\newcommand{\PWfL}{\Gamma_{\ZtofL}}
\newcommand{\FWZ}{\Gamma_{\PZ}}
\newcommand{\BFLL}{\Br(\ZtoLL)}
\newcommand{\BF}{\BFfL}
\newcommand{\BFfL}{\Br(\ZtofL)}
\newcommand{\BFfM}{\Br(\ZtofM\!)}
\newcommand{\BFtMtE}{\Br(\ZtotMtE\!)}
\newcommand{\BFfE}{\Br(\ZtofE\!)}
\newcommand{\Zo}{\PZ_1}
\newcommand{\Zt}{\PZ_2}
\newcommand{\sixL}{6\Pell}
\newcommand{\ZtosixL}{\PZ \to \sixL}
\newcommand{\BFsixL}{\Br({\ZtosixL})}
\newcommand{\dxy}{\abs{d_{xy}}}
\newcommand{\dz}{d_z}
\newcommand{\IP}{\text{IP}}
\newcommand{\RelPFIso}{I_\text{rel}}
\newcommand{\Aeff}{A_\text{eff}}
\newcommand{\mfL}{m_{\fL}}
\newcommand{\mZn}[1]{m_{\PZ_{#1}}}
\newcommand{\mZo}{\mZn{1}}
\newcommand{\mZt}{\mZn{2}}
\newcommand{\ptL}{\pt^{\Pell}}
\newcommand{\ptLn}[1]{\pt^{\Pell_{#1}}}
\newcommand{\ptLo}{\ptLn{1}}
\newcommand{\ptLt}{\ptLn{2}}
\newcommand{\etaL}{\eta^{\Pell}}
\newcommand{\mLL}{m_{\LL}}
\newcommand{\TPO}{\sgn(\sin \phi)}
\newcommand{\Asinphi}{A_{\sin\phi}}
\newcommand{\dGdx}{\ddinline{\PWfL}{x}}
\newcommand{\dGd}[1]{\ddinline{\PWfL}{#1}}
\newcommand{\pLo}{p_{\Pell_1}}
\newcommand{\mtrlttt}{m_{\Pell_{2,3,4}}}
\newcommand{\sinphi}{\sin\phi}
\newcommand{\NP}{\text{NP}}
\newcommand{\SM}{\text{SM}}
\newcommand{\MadGraph}{\MADGRAPH}
\newcommand{\FeynRules}{\textsc{FeynRules}\xspace}
\newcommand{\NNPDF}{\text{NNPDF}\xspace}

\newlength{\cmsFigWidth}
\setlength{\cmsFigWidth}{0.48\textwidth}

\title{Studies of \texorpdfstring{$\PZ \to 4\Pell$}{Z to 4L} decays in proton-proton collisions at \texorpdfstring{$\sqrt{s} = 8$ and $13\TeV$}{sqrt(s) = 8 and 13 TeV}}

\date{\today}

\abstract{
Decays of $\PZ$ bosons to four charged leptons (electrons and muons) are studied in proton-proton collisions at $\sqrt{s} = 8$ and 13\TeV. The analysis is based on data collected with the CMS detector at the LHC corresponding to an integrated luminosity of 19.7\fbinv at 8\TeV and 138\fbinv at 13\TeV. The measured value of the inclusive branching fraction for all four-lepton decay modes, $\BF$, is $[4.67 \pm 0.11 \stat \pm 0.10 \syst] \times 10^{-6} $, which has a precision of about 3\% limited by both statistical and systematic uncertainties. Measurements of the individual branching fractions for the decays $\PZ \to 4\Pgm$, $\PZ \to 4\Pe$, and $\PZ \to 2\Pgm 2\Pe$ are also reported. Differential decay rates are presented as functions of kinematic and angular quantities in the $\PZ$ boson rest frame. Measurements of triple-product asymmetries, which are sensitive to possible violations of charge conjugation and parity invariance, are performed for $\PZ \to 4\Pell$ decays. The results are compared with standard model predictions and used to set limits on the production of new gauge bosons.
}

\hypersetup{
pdfauthor={CMS Collaboration},
pdftitle={Studies of \texorpdfstring{$\PZ \to 4\Pell$}{Z to 4L} decays in proton-proton collisions at \texorpdfstring{$\sqrt{s} = 8$ and $13\TeV$}{sqrt(s) = 8 and 13 TeV}},
pdfsubject={CMS},
pdfkeywords={CMS, standard model, rare decays}}

\maketitle

\section{Introduction} \label{sec:Introduction}
\begin{figure}[hb] \centering
\includegraphics{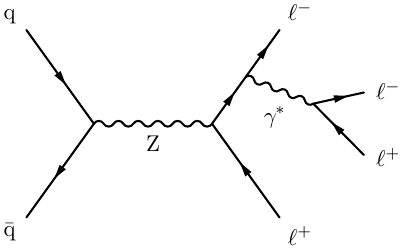}
\caption{The $\ZtofL$ decay process in $\PP$ collisions.}
\label{fig:FeynmanDiagram}
\end{figure}

Decays of on-shell $\PZ$ bosons to four-lepton final states $\ZtofL$ (where $\Pell = \Pe, \Pgm$)
are predicted by the standard model~(SM) to be rare, with a branching fraction $\BF$ on the order of~$10^{-6}$.
The Feynman diagram in Fig.~\ref{fig:FeynmanDiagram} represents the $\ZtofL$ decay process in the SM.
These decays provide unique opportunities to observe physics beyond the SM (BSM), but because
they are so rare, they have not yet been studied in detail.
New light gauge bosons, if they exist, might be observable in $\ZtofL$ decays.
For example, bosons that couple predominantly to leptons
will contribute to the $4\Pell$ final state leading to a measured value of $\BF$, which differs
from the SM predicted value.

The high integrated luminosities recorded at the CERN LHC enable precise measurements of the $\ZtofL$ decay process.
Previous measurements of $\BF$ have been made by the CMS Collaboration using proton-proton (\PP) collision data
collected at $\sqrt{s}=7\TeV$~\cite{CMS12} and 13\TeV~\cite{CMS16, CMS18}, and by the ATLAS Collaboration using
combined 7 and 8\TeV data sets~\cite{ATLAS14} and 13\TeV data~\cite{ATLAS19,ATLAS21}.
All of these measurements agree with the SM value, $\BF = (4.70 \pm 0.03)\ten{$-6$}$, obtained using \MADGRAPH~\cite{MadGraph}.
Recently, CMS published a search for \PZ~decays to $\PGt\Pgt\PGm\PGm$~\cite{CMS:eetautau}.
The LHCb Collaboration published an observation of the decay $\JPsi \to \PGmp\PGmm\PGmp\PGmm$~\cite{LHCb:Jpsi4mu}
as did CMS using a method~\cite{CMS:Jpsi4mu} similar to the one described here.

The presence of a new light boson with nonzero couplings to electrons and/or muons would lead to an increase in $\BF$.
Measurements of $\BF$ at the LHC have been used to set limits on the masses of BSM bosons with lepton couplings~\cite{Rainbolt,Drees2018}
including the light gauge boson appearing in certain leptophilic dark matter models~\cite{LeptophilicLimits}
and models based on an $L_\mu-L_\tau$ $U(1)$ local gauge symmetry.
CMS performed a direct search for an $L_\mu-L_\tau$ gauge boson in the $\ZtofM$ channel
using data corresponding to an integrated luminosity of 77.3\fbinv collected at 13\TeV~\cite{EXO}.
ATLAS published a similar search for $\PZpr$ bosons decaying to $\MM$~\cite{ATLASexo}.
More precise and detailed studies of $\ZtofL$ decays will improve the understanding of this process
potentially leading to improved sensitivity to BSM physics.

We present an in-depth study of the $\ZtofL$ process using CMS \PP collision data collected at 8~and 13\TeV.
The analyzed data samples correspond to integrated luminosities of $\lumi = 19.7\fbinv$ at 8\TeV recorded in 2012 and
of 138\fbinv at 13\TeV in 2016--2018~\cite{CMS-LUM-17-003,CMS-PAS-LUM-17-004,CMS-PAS-LUM-18-002}.
We report the most precise measurement of $\BF$ to date and the first CMS measurements of the branching fractions
for the distinct $\fM$, $\fE$, and $\tMtE$ channels. Measurements of differential decay rates for the $\ZtofL$ process
as functions of several kinematic and angular quantities defined in the \PZ boson rest frame are presented as well.
The ATLAS Collaboration has published similar measurements of differential cross sections~\cite{ATLAS21}.
We also report measurements of triple-product asymmetries that may be used to test for violations of
charge conjugation and parity (\CP) invariance in $\ZtofL$ decay.
Our measurements are compared with SM predictions, and constraints are derived on a BSM boson lighter than the \PZ boson
and with nonzero couplings to electrons and muons.

This paper is structured as follows. The analysis strategy is outlined in Section~\ref{sec:Strategy} and
the CMS detector is described in Section~\ref{sec:Detector}, highlighting the components relevant to this analysis.
Section~\ref{sec:Samples} provides an account of the data and simulated event samples used in the analysis.
The object and event reconstruction and selection criteria are described in Section~\ref{sec:Selection}, and
our methods for estimating and accounting for background contributions are explained in Section~\ref{sec:Background}.
Section~\ref{sec:Systematics} describes the sources of systematic uncertainty and the methods
used for their evaluation. The results are presented and interpreted in Section~\ref{sec:Results}.
A summary highlighting the main findings of the analysis is given in Section~\ref{sec:Summary}.

\section{Analysis strategy} \label{sec:Strategy}
The LHC and the CMS detector serve as the source of a very large number of
\PZ\,bosons---far more than produced at the Large Electron-Positron Collider, LEP---enabling
the measurement of the branching fraction $\BFfL$ as well as differential decay rates.
However, $\BFsixL \approx 10^{-10}$, which remains beyond the sensitivity of current CMS data.

The physics quantities of interest are defined in the rest frame of the \PZ~boson and are
practically independent of the center-of-mass energy. The phase space of
the $\ZtofL$ decay process, with respect to which theoretical predictions of branching fractions
and differential decay rates are calculated, is defined by two requirements. First, the
four-lepton invariant mass must satisfy $80 < \mfL < 100\GeV$. Second, any pair
of same-flavor, opposite-sign leptons must satisfy $\mLL > 4\GeV$. This second requirement
removes contributions from low-mass resonances such as $\JPsi \to \Pellp\Pellm$,
which has been observed in \PZ\,boson decays~\cite{CMSJpsi}.
Theoretical predictions of branching fractions and differential decay rates are calculated
with respect to this phase space. When working
with simulated event samples, generator-level leptons are ``dressed'' by summing
all photons falling within a cone of $\Delta R = \sqrt{\smash[b]{(\Delta\eta)^2 + (\Delta\phi)^2}} = 0.1$
centered on the charged lepton, where $\Delta\eta$ is the difference in pseudorapidity and $\Delta\phi$ is
the difference in azimuthal angle measured in radians.
The $\ZtoLL$ decay process is used for normalizing $\BF$, as explained below,
and its phase space is defined simply by the requirement $80 < \mLL < 100\GeV$.
The phase space should not be confused with the fiducial region defined by
the event selection requirements stated in Section~\ref{sec:Events}.

The number of selected signal events is
\begin{equation} \label{eqn:NSignalEvents}
  \Nsig = \BF \frac{{\axe_{\fL}}}{(1-\fnr)} \, \sigma(\PP\to\PZ+X) \lumi,
\end{equation}
where \lumi is the integrated luminosity,
$A$ is the acceptance for events in the phase space region to fall in the
fiducial region, $\epsilon$ is the efficiency for selecting events in the fiducial region
that pass the event selection requirements, $\fnr$ is a small correction
for nonresonant production of $\fL$ events, and $\sigma(\PP\to\PZ+X)$ is the cross
section for the inclusive production of \PZ bosons.

The branching fraction $\BFfL$ could be extracted directly from Eq.~(\ref{eqn:NSignalEvents}).
Significant uncertainties, however, would ensue from the measured values of the cross section and integrated luminosity.
In order to eliminate these uncertainties, the production of lepton pairs,
$\PP\to\PZ\to\LL$, is used as a precise measure of $\sigma(\PP\to\PZ+X) \lumi$.
Since
\begin{equation*} \label{eqn:NDileptonEvents}
  N_{\LL} = \BFLL {\axe_{\LL}} \, \sigma(\PP\to\PZ+X) \lumi ,
\end{equation*}
the percent-level uncertainties of the quantity $\sigma(\PP\to\PZ+X) \lumi$ are exchanged
for extremely small statistical uncertainties for $N_{\LL}$ and the well-measured value
for $\BFLL$~\cite{PDG24}. By using similar fiducial regions for $\ZtofL$ and $\ZtoLL$,
the relative uncertainty in $\axe_{\fL}$ and $\axe_{\LL}$ are also rather small.
On this basis, we infer $\BF$ using the formula
\ifthenelse{\boolean{cms@external}}
{
\begin{equation}
\label{eqn:BranchingFraction}
\begin{aligned}
    \BF &= \BFLL (1 - \fnr) \\
        &\times \, \frac{\axe_{\LL}}{\axe_{\fL}} \frac{(\Nobs - \Nbkg)_{\fL}}{(\Nobs - \Nbkg)_{\LL}} .
\end{aligned}
\end{equation}
}
{
\begin{equation}
\label{eqn:BranchingFraction}
    \BF = \BFLL (1 - \fnr) \frac{\axe_{\LL}}{\axe_{\fL}} \frac{(\Nobs - \Nbkg)_{\fL}}{(\Nobs - \Nbkg)_{\LL}} .
\end{equation}
}
In this expression, $\BFLL = (3.3658 \pm 0.0023)\%$ is the world-average dilepton branching fraction~\cite{PDG24}.
The quantity $\fnr = 0.04 \pm 0.01$ is the fraction~\cite{CMS18} of $\fL$ events from nonresonant processes
such as $\ZZtofL$ and is calculated using \MGvATNLO~v.3.5.0~\cite{MadGraph}.
The quantity $\axe$ is the product of theoretical acceptance and experimental efficiency for the given final state.
The $\Nobs$ is the observed number of selected events and $\Nbkg$ is the estimated number of background events.
These quantities are defined separately for the $\fL$ signal channel and the $\LL$ normalization channel.
The formulation of Eq.~(\ref{eqn:BranchingFraction}) is founded on two ratios of similar quantities---$\axe$
for $\LL$ and $\fL$, and the ratio of background-corrected yields---and consequently is experimentally
robust. The constants in the formula, $\BFLL$ and $(1-\fnr)$, are precisely known.

The flavor-specific branching fractions $\BFfM$, $\BFfE$, and $\BFtMtE$ are measured using a
choice of normalization channels that reduces the systematic uncertainties coming from lepton
reconstruction and identification. For $\BFfM$, only $\MM$ events are used in
Eq.~(\ref{eqn:BranchingFraction}), and for $\BFfE$ only $\EE$ events are used.
With this choice,
\begin{equation*} \label{eqn:BF4M}
\BFfM \propto \left. \frac{(\Nobs - \Nbkg)_{\fM}}{\axe_{\fM}} \middle/ \frac{(\Nobs - \Nbkg)_{\MM}}{\axe_{\MM}} \right.
\end{equation*}
and
\begin{equation*}  \label{eqn:BF4E}
\BFfE \propto \left. \frac{(\Nobs - \Nbkg)_{\fE}}{\axe_{\fE}} \middle/ \frac{(\Nobs - \Nbkg)_{\EE}}{\axe_{\EE}} \right..
\end{equation*}
Both $\MM$ and $\EE$ events are used to measure $\BFtMtE$:
\ifthenelse{\boolean{cms@external}}
{
\begin{equation*}  \label{eqn:BF2M2E}
\BFtMtE \propto
 \frac{ 2 \frac{(\Nobs - \Nbkg)_{\tMtE}}{\axe_{\tMtE}} }
      { \Biggl(\frac{(\Nobs - \Nbkg)_{\MM}}{\axe_{\MM}} + \frac{(\Nobs - \Nbkg)_{\EE}}{\axe_{\EE}} \Biggr) }.
\end{equation*}
}
{
\begin{equation*}  \label{eqn:BF2M2E}
\BFtMtE \propto 2 \, \frac{(\Nobs - \Nbkg)_{\tMtE}}{\axe_{\tMtE}} \left/ \middle(\frac{(\Nobs - \Nbkg)_{\MM}}{\axe_{\MM}} + \frac{(\Nobs - \Nbkg)_{\EE}}{\axe_{\EE}} \right).
\end{equation*}
}
The inclusive branching fraction $\BF$ is defined to be the sum of the measurements of flavor-specific final states:
\begin{equation*}  \label{eqn:BF4L}
\BF \equiv \BFfM + \BFtMtE + \BFfE.
\end{equation*}

\subsection{Differential decay rates}

Differential decay rates, $\ddinline{\PWfL}{x}$, where $\PWfL$ is the partial width for $\BF$ and
$x$ is one of nine kinematic quantities listed below, are defined in the \PZ\,boson rest frame,
which is well-approximated by the four-lepton rest frame.
For the sake of clearly defining the differential decay rates,
we label the more~(less) massive same-flavor, opposite-sign lepton pair in the event as $\Zo$~($\Zt$).
This notation does not imply that we necessarily view $\ZtofL$ as a $1 \to 2$ decay process.
Furthermore, we order the momenta of the four leptons in the \PZ\,boson rest frame, so that
$p_{\Pell_1} > p_{\Pell_2} > p_{\Pell_3} > p_{\Pell_4}$; in general,
this will not correspond to the transverse momentum (\pt) ordering of leptons, and the leading
lepton in the \PZ\,boson rest frame is not necessarily the lepton with highest\,\pt.
For the differential distributions, $x$ will refer to one of the following kinematic quantities:
\begin{itemize}
\item $\mZo$~($\mZt$), the invariant mass of the $\Zo$~($\Zt$) lepton pair;
\item $\pLo$, the momentum of the leading lepton in the $\PZ$ rest frame;
\item $\mtrlttt$, the invariant mass of the three trailing leptons;
\item $\alpha_{\Zo}$~($\alpha_{\Zt}$), the angle between the momentum vectors of the leptons of pair $\Zo$~($\Zt$);
\item $\beta$, the angle between the momentum vector of the less-energetic lepton of the $\Zo$ pair and the momentum vector of $\Zt$;
\item $\theta_{\Zo}$~($\theta_{\Zt}$), the angle between the momentum vector of the positively-charged lepton of pair $\Zo$~($\Zt$) and the momentum vector of $\Zt$~($\Zo$), measured in the rest frame of $\Zo$~($\Zt$).
\end{itemize}

\subsection{A probe for the violation of \texorpdfstring{\CP}{CP} invariance} \label{sec:CPViolationProcedure}

We probe for a violation of \CP invariance in $\ZtofL$ decays by measuring a triple-product asymmetry.
In the SM, $\ZtofL$ decays are expected to conserve \CP.
References~\cite{Bevan,Bevan2} suggest a probe of \CP invariance for the $\ZtofL$ decay process, and define the observable
$\phi$ as the angle between the decay planes of $\Zo$ and $\Zt$ in the $\PZ$ boson rest frame.
The ``decay plane'' of a pair is defined by the momentum vectors of its constituent leptons,
which in general are not collinear.
The angle $\phi$ is measured from the ``front'' face of the $\Zo$ plane to the ``front'' face of
the $\Zt$ plane, where the ``front`` of the $\Zo$ plane is defined by the outward-pointing unit normal vector
\begin{equation*}
\hat{n}_{\Zo} =
\frac{ \hat{p}_{\Zo}^{\Pellp} \times \hat{p}_{\Zo}^{\Pellm} }
     { \lvert \hat{p}_{\Zo}^{\Pellp} \times \hat{p}_{\Zo}^{\Pellm} \rvert }
\end{equation*}
and similarly for the $\Zt$ plane. Here, $\hat{p}_{\Zo}^{\Pellp}$ is the unit vector in the direction
of the momentum of the positively charged $\Zo$ lepton. It is convenient to define $\phi$ as
\begin{equation*} \label{eqn:SinPhiDefinition}
   \sin\phi = (\hat{n}_{\Zo} \times \hat{n}_{\Zt}) \cdot \hat{p}_{\Zo},
\end{equation*}
where $\hat{p}_{\Zo}$ is the unit vector in the direction of the momentum of $\Zo$.

The quantity $\TPO$, where $\sgn$ is the sign function,
is a triple-product asymmetry observable that may be used to test for {$\mathit{C}$}, {$\mathit{P}$}, and \CP asymmetries.
It is equivalent to $\sgn(\psi)$, where $\psi = (\vec{p}_{\Zo}^{\,\Pellp} \times \vec{p}_{\Zo}^{\,\Pellm}) \cdot \vec{p}_{\Zt}^{\,\Pellp}$
is a scalar triple-product uniquely defined by the four lepton momenta in the $\PZ$ boson rest frame~\cite{Bevan}.
The asymmetry $\Asinphi$ is given by
\begin{equation} \label{eqn:TripleProductAsymmetry}
    \Asinphi = \frac{N_+ - N_-}{N_+ + N_-},
\end{equation}
where $N_+$~($N_-$) is the number of events for which $\sin\phi > 0$\,($< 0$) after subtracting expected
background contributions. The distribution of $\sin\phi$ might itself show evidence of an asymmetry,
which would constitute a contradiction of SM expectations.
Similar probes have been applied to four-body meson decays such as
$\PDz \to \PKp \PKm \Pgpp \Pgpm$ studied by the Belle Collaboration~\cite{Belle}
and $\PKzLS \to \Pgpp \Pgpm \EE$ studied by the NA48 Collaboration~\cite{NA48}.

\subsection{Physics beyond the SM} \label{sec:LimitsProcedure}

\begin{figure} \centering
\includegraphics{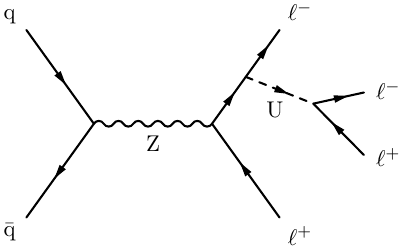} \hfill
\includegraphics{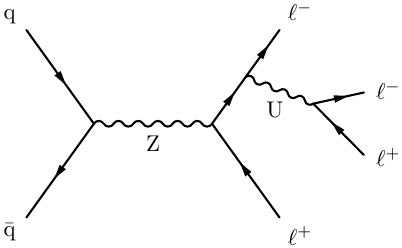}
\caption{The $\ZtofL$ process mediated by a scalar (left) or vector (right) boson $\PU$.}
\label{fig:NPFeynmanDiagrams}
\end{figure}

We consider possible contributions from a new neutral scalar or vector boson $\PU$ which may mediate
the $\ZtofL$ decay, as shown in Fig.~\ref{fig:NPFeynmanDiagrams}. This $\PU$ boson couples to electrons
and muons, potentially with different strengths. Couplings to other fermions and to gauge bosons are not considered.
The scalar and vector models, respectively, are described by the following additions to the SM Lagrangian
\begin{equation} \label{eqn:ScalarLagrangian}
   \mathcal{L} \supset \frac{1}{2} \partial_\mu \PU \partial^\mu \PU - \frac{1}{2} m_{\PU}^2 \PU^2 + g_{\Pell} \bar{\Pell} \Pell \PU
\end{equation}
and
\begin{equation} \label{eqn:VectorLagrangian}
   \mathcal{L} \supset \frac{1}{2} \partial_\mu \PU_\nu \partial^\mu \PU^\nu
   + \frac{1}{2} m_{\PU}^2 \PU_\mu \PU^\mu + g_{\Pell} \bar{\Pell} \gamma_\mu \Pell \PU^\mu,
\end{equation}
where $m_\PU$ is the mass of the $\PU$ boson and $g_{\Pell}$~($= g_{\Pe}, g_{\PGm}$) is the strength
of its couplings to leptons.

The combined value of published $\BF$ measurements is used in Ref.~\cite{Rainbolt} to set limits on
the production of $\PU$ bosons. A similar approach is taken in Ref.~\cite{Drees2018}.
The flavor-specific branching fractions $\BFfM$ and $\BFfE$ enable us
to set limits on the presence of $\PU$, a new neutral scalar or vector boson
which couples only to muons~(\ie, with $g_{\Pe} = 0$)
or only to electrons~($g_{\PGm} = 0$). Our $\BFtMtE$ measurement could be used to constrain the case
where $g_{\Pe}, g_{\PGm} > 0$ but $g_{\Pe} \neq g_{\PGm}$, although this case is not investigated
in this paper.

\section{The CMS detector and event reconstruction} \label{sec:Detector}
The central feature of the CMS apparatus is a superconducting solenoid of 6\unit{m} internal diameter, providing a magnetic field of 3.8\unit{T}. Within the solenoid volume are a silicon pixel and strip tracker, a lead tungstate crystal electromagnetic calorimeter (ECAL), and a brass and scintillator hadron calorimeter (HCAL), each composed of a barrel and two endcap sections. Forward calorimeters extend the $\eta$ coverage provided by the barrel and endcap detectors. Muons are detected in gas-ionization detectors embedded in the steel flux-return yoke outside the solenoid.
A more detailed description of the CMS detector, along with a definition of the coordinate system used and
relevant kinematic variables, is reported in Ref.~\cite{Chatrchyan:2008zzk}.

The particle-flow (PF) algorithm~\cite{CMS-PRF-14-001} uniquely identifies all reconstructed particles in an event
based on an optimized combination of information from many different elements of the CMS detector. PF objects include
charged and neutral hadrons, photons, electrons, and muons.

Muons are identified~\cite{muID} as tracks in the central tracker consistent with either a track or several hits in the
muon system, and with energy deposits consistent with the expectation for a muon traversing the calorimeters.
Muons are detected in the range $\abs{\eta} < 2.4$ using detection planes employing
three technologies: drift tubes, cathode strip chambers, and resistive plate chambers.
For muons with \pt up to 100\GeV, the \pt resolution is determined mainly by the tracker, and amounts
to 1\% in the barrel and 3\% in the endcaps~\cite{Chatrchyan:2012xi, muID}.

An electron is identified~\cite{eID,eID8} in the range $\abs{\eta}< 2.5$
as a charged particle track and one or more ECAL energy clusters consistent with the
extrapolation of this track to the ECAL surface and with bremsstrahlung photons emitted as the electron propagates
through the tracker material. The momentum of the electron is estimated by combining the ECAL cluster energy measurements
with the momentum measurement in the tracker. The momentum resolution for electrons with $\pt \approx 45\GeV$
ranges 1.7--4.5\%. It is generally better in the barrel region than in the endcaps, and depends on the
amount of energy emitted by the electron as bremsstrahlung~\cite{eID8}.

The energy of photons is obtained from the measurement of ECAL clusters. The energy of charged hadrons
is determined from a combination of their momentum as measured in the tracker and the matching ECAL and HCAL
energy deposits.
The energy of neutral hadrons is obtained from ECAL and HCAL cluster energies after
the deposits from charged hadrons, photons, electrons, and muons have been subtracted.

The imbalance of the total \pt, known as missing \pt, is computed as the opposite
of the vector sum of the \ptvec of all charged and neutral PF candidates~\cite{MET}. The impact of additional $\PP$ collisions
in the same or adjacent bunch crossings (``pileup'') is handled as described in Refs.~\cite{JME-10-003,PILEUP}.

The reconstructed vertex with the largest sum of $\pt^2$ of all physics objects associated to that vertex
is taken to be the primary \PP interaction vertex in the event. These physics objects may include jets reconstructed
using the anti-\kt clustering algorithm~\cite{Cacciari:2008gp,Cacciari:2011ma} taking tracks associated with the
given vertex and the associated missing momentum vector as input.

\section{Data and simulated event samples} \label{sec:Samples}
The results of this analysis are based on the CMS data sets collected at $\sqrt{s} = 8$ and 13\TeV.
The data are collected using triggers that require the presence of one muon candidate, one electron candidate,
or two same-flavor lepton candidates~\cite{TRG12,TRG17}.
The trigger path \pt requirements vary from one data-taking period to another.
The single-muon~(single-electron) paths used in this analysis require a candidate lepton with
$\pt^{\PGm} > 24$--27\GeV~($\pt^{\Pe} > 27$--35\GeV). The dimuon~(dielectron) paths require one candidate with
$\pt^{\PGm_1} > 17$--23\GeV~($\pt^{\Pe_1} > 17$--23\GeV) and another one with
$\pt^{\PGm_2} > 8\GeV$\,($\pt^{\Pe_2} > 8$--12\GeV).

Samples of simulated signal and background events are used to optimize the event selection criteria,
evaluate the detector acceptance and efficiency of each selection, verify the procedure for
a background estimate based on control samples in data, and assess systematic uncertainties.
The available Monte Carlo (MC) event generators vary somewhat from one data-taking period to the next.
Most simulated event samples are generated at next-to-leading order~(NLO) in perturbative quantum chromodynamics (QCD)
for \PP collisions at $\sqrt{s} = 8\TeV$ and 13\TeV. However, some 8\TeV samples were generated at leading order (LO).

We begin by describing the 13\TeV simulated samples.
The $\ZtofL$ signal process is simulated at NLO using the event generators {\POWHEG}~2.0~\cite{Nason:2004rx, Frixione:2007vw, Alioli:2010xd, Melia:2011tj}
and \MGvATNLO~v.2.4.2~\cite{MadGraph}. The results obtained with these two event generators are compatible.
The $\ZtoLL$ normalization process is simulated using the \MGvATNLO~v.2.2.2 (for 13\TeV) and
the LO generator \MADGRAPH5 (for 8\TeV) event generators, respectively.

Diboson $\WW$ events are generated with \POWHEG, $\WZ$ with \MGvATNLO, and $\ZZ$ with both \MGvATNLO and \POWHEG.
Top quark production processes $\ttbar(+\PZ)$
are simulated using \MGvATNLO and \POWHEG. Triboson processes
are simulated using \MGvATNLO. Finally, Higgs boson production processes with $\PH \to \ZZ$
are simulated with \MGvATNLO.

The 8\TeV simulated samples are simulated in ways similar to the 13\TeV samples. The $\ZtofL$ ($\ZtoLL$) process is
simulated using the \POWHEG (\MADGRAPH) generator, respectively. Diboson processes are simulated at LO with \MADGRAPH
as are $\ttbar(+\PZ)$ processes. Due to small cross sections, triboson and Higgs boson processes
contribute negligible backgrounds at 8\TeV and are not simulated.

Parton distribution functions (PDFs) are essential for event generators and cross section calculations.
The simulations of 2016 (2017, 2018) data rely on the NNPDF\,3.0~\cite{NNPDF30} (NNPDF\,3.1~\cite{NNPDF31})
PDF sets and the simulations of 2012 8\TeV data rely on the CT10~\cite{CT10} and NNPDF\,2.3~\cite{NNPDF23}
PDF sets.

For these simulated samples, parton showering, hadronization, and multiparton interactions are modeled
using \PYTHIA 6.426, 8.226, and 8.230~\cite{Sjostrand:2000wi, Sjostrand:2014zea}.
Various so-called ``tunes'' (parameter sets) are used to improve the simulation of the underlying event which
includes multiparton interactions and other soft effects of the $\PP$ collision.
For the 2016 13\TeV simulated samples, the CUETP8M1 tune~\cite{CUETP8M1}
is used, although the CUETP8M2T4 tune~\cite{CUETP8M2T4} is used for top quark processes.
For the 2017 and 2018 samples, the CP5 tune~\cite{CP5} is used, and for the 8\TeV samples,
the Z2$^*$ tune~\cite{Z2STAR} is used.

The CMS detector response is simulated with the \GEANTfour package~\cite{Agostinelli:2002hh,GEANT4B}.
The simulation of in-time and out-of-time pileup is calibrated to match the distribution of pileup in data~\cite{PILEUP}.
The average number of \PP collisions in an event is about 20 in the 8\TeV data and ranges
23--32 in the 13\TeV data, depending on the data-taking period.

Samples of simulated events including BSM contributions to the $\ZtofL$ process are used to set limits on the
parameters of new physics (NP) models including BSM scalar and vector bosons. The implementation follows
Ref.~\cite{Rainbolt} and is carried out using the \FeynRules package~\cite{FeynRules}.
Event samples are generated using \MADGRAPH~\cite{MadGraph} at representative values of model parameters.
These samples include contributions from the SM $\ZtofL$ process, and are generated at 13\TeV to LO.
Acceptance differences for events at 13\TeV and 8\TeV have a negligible effect on the results.
For these samples, parton showering and detector simulation are not performed
because the analysis depends on reconstructed lepton quantities.

\section{Object and event selection} \label{sec:Selection}
The lepton identification used in these studies is based on
Refs.~\cite{eID8,eID,Chatrchyan:2012xi,muID}.
The event selection involves only kinematic quantities based on either two or four well-reconstructed leptons.

\subsection{Object selection} \label{sec:Objects}

Electrons and muons are selected on the basis of kinematic, impact parameter, and reconstruction criteria.
A first set of criteria are referred to as the ``loose'' identification requirements, and a second, more restrictive
one is described in the next paragraph.
A loosely-identified electron~(muon) candidate must satisfy $\pt > 7$\,(5)\GeV and $\abs{\eta} < 2.5$\,(2.4).
A candidate of either flavor must satisfy the transverse and longitudinal impact parameter requirements $\dxy < 0.5\cm$ and
$\abs{\dz} < 1\cm$ with respect to the event's primary vertex. The three-dimensional impact parameter
($\IP$) of each candidate must also satisfy the requirement $\abs{\IP / \sigma_{\IP}} < 4$
with respect to the primary vertex, where $\sigma_{\IP}$ is the uncertainty in the $\IP$ measurement.

Further ``tight'' identification requirements are defined that select leptons originating from the hard process of an event
and suppress leptons resulting from decays of hadrons containing a \PQb or \PQc quark;
these requirements are applied to lepton candidates that pass
the loose identification criteria. Tightly-identified muons must be identified as such by the PF algorithm described in
Section~\ref{sec:Detector}, and the tight identification requirement for electrons is defined by a multivariate classifier.
Each tightly-identified lepton must be isolated in the sense that the sum of energies of particles close to the candidate
lepton is small compared with the lepton energy. Leptons emitted in \PZ~boson decays are generally expected to be
isolated, in contrast to ``nonprompt'' leptons emitted in heavy hadron decays that are found inside jets;
the isolation requirement suppresses unwanted nonprompt leptons. The relative isolation of a lepton is defined by
\ifthenelse{\boolean{cms@external}}
{
\begin{equation*} \label{eqn:Isolation}
\begin{aligned}
\RelPFIso^{\Pell} & = \Bigl(\sum \pt^\text{charged hadrons} \\
  & + \max\left[\sum \pt^\text{neutral hadrons} - \pt^\text{pileup}, 0 \right] \Bigr) \Bigl/ \pt^{\Pell} ,
\end{aligned}
\end{equation*}
}{
\begin{equation*} \label{eqn:Isolation}
\RelPFIso^{\Pell} = \left. \left(\sum \pt^\text{charged hadrons} + \max\!\left[\sum \pt^\text{neutral hadrons} - \pt^\text{pileup}, 0 \right]\right) \middle/ \pt^{\Pell} \right. ,
\end{equation*}
}
where the sum is over PF objects within a cone of size $\DR$ about the lepton's direction.
A small value of $\RelPFIso$ indicates an isolated lepton. For events collected at 8\,(13)\TeV, $\DR = 0.4$\,($0.3$).
For muons, $\pt^\text{pileup} = \sum_\text{PU} \pt^\text{charged hadrons}/2 - \sum \et^\text{photons}$
where $\text{PU}$ represents the pileup vertices, and the factor of $1/2$ is motivated by the fact that
two charged hadrons are expected for each neutral one. We require $\RelPFIso^{\PGm} < 0.4$\,($< 0.35$) at 8\,(13)\TeV.
For electrons, $\pt^\text{pileup} = \rho\Aeff$, where $\rho$ is the mean energy density in the event and the
effective area $\Aeff$ is the ratio of the slope of the average isolation and the slope of $\rho$
as functions of the number of pileup vertices~\cite{eSubtraction}. We require $\RelPFIso^{\Pe} < 0.4$ at 8\TeV.
The multivariate classifier applied to the 13\TeV data samples includes the object relative isolation as an input variable,
so we impose no explicit isolation requirement on electrons in those samples.

The energies and momenta of leptons are corrected on a per-lepton basis for any bias related to detector misalignment or calibration.
In simulated events, the object energies are further smeared such that the energy distributions for leptons of both flavors match
those observed in data. Unless otherwise specified, all leptons mentioned in the following pass the tight identification requirements.

\subsection{Event selection} \label{sec:Events}

A set of kinematic and topological requirements forms the basis of the event selection
and also defines the fiducial region for acceptance in the CMS detector.
These requirements are stated below and summarized in Table~\ref{tab:FiducialRegion}.
In order to be selected, an event must satisfy the requirements for both the fiducial and
the phase space regions. The phase space is defined above, in Section~\ref{sec:Strategy}.
The fiducial regions are as similar as possible for the $\ZtofL$ and $\ZtoLL$ final states:
$\abs{\eta^{\Pell}} < 2.5$ for all leptons, and $\ptLo > 20\GeV$ and $\ptLt > 10\GeV$ for
the leading and subleading leptons, respectively.
The trailing two leptons of a four-lepton event are both required to have $\ptL > 5\GeV$.

\begin{table}
\centering
\topcaption{Definitions of the phase space and fiducial regions for four-lepton and for dilepton events.
The $\mLL$ requirement for four-lepton events applies to all possible same-flavor, opposite-sign lepton pairs.}
\begin{scotch}{l l l}
Region  & $\fL$ events & $\LL$ events \\
\hline
Phase space & $80 < \mfL < 100\GeV$ & $80 < \mLL < 100\GeV$ \\
            & $\mLL > 4\GeV$ & \NA \\[\cmsTabSkip]
Fiducial & $\abs{\eta^{\Pell}} < 2.5$ & $\abs{\eta^{\Pell}} < 2.5$ \\
         & $\ptLo > 20\GeV$ & $\ptLo > 20\GeV$ \\
         & $\ptLt > 10\GeV$ & $\ptLt > 10\GeV$ \\
         & $\ptLn{3}, \ptLn{4} > 5\GeV$ & \NA  \\
\end{scotch}
\label{tab:FiducialRegion}
\end{table}

A four-lepton event must contain exactly four tightly-identified leptons and fall within the
four-lepton fiducial and phase space regions. The four leptons must have flavor and charge
such that they can be arranged as two pairs of same flavor and opposite sign.
These pairs are denoted $\Zo$ and $\Zt$, where $m_{\Zo} > m_{\Zt}$. With the SM in mind,
$\Zt$ is the lepton pair produced mainly via the virtual photon emission shown in Fig.~\ref{fig:FeynmanDiagram}.
The pairing is ambiguous for events with four leptons of the same flavor; this ambiguity is resolved by testing all possible
configurations of opposite-sign pairs and choosing $\Zo$ and $\Zt$ to maximize the difference $m_{\Zo} - m_{\Zt}$.
We require $m_{\Zo} > 12\GeV$ in order to reduce background from $\PgU$~resonances, which can be difficult to
estimate accurately. Finally, all same-flavor leptons within the event must be separated from one another
by $\DR > 0.02$, and all electrons and muons must be separated from one another by $\DR > 0.05$.
These separation requirements remove some poorly reconstructed leptons and reduce contributions
from bremsstrahlung photons that convert in the detector. They have a negligible impact on the
acceptance of $\ZtofL$ decays.

A dilepton event is selected if it falls within the appropriate fiducial region and contains exactly two leptons
satisfying the tight identification criteria. The two leptons must be of the same flavor and of opposite charge
and must be separated from one another by $\DR > 0.02$.

The event-level combined lepton selection efficiencies are 37\% for $\ZtofE$, 70\% for $\ZtofM$,
and 51\% for $\ZtotMtE$, and for dilepton channels, they are 72\% for $\ZtoEE$ to 91\% for $\ZtoMM$.

\section{Background estimation} \label{sec:Background}
The largest sources of background events come from the $\ZtoLL$ and $\ttbar \to 2\Pell 2\PGn$ production processes.
In both cases, the $\fL$ requirements outlined in Section~\ref{sec:Events} select the two prompt leptons and two
additional, nonprompt leptons despite the isolation requirements stated in Section~\ref{sec:Selection}.
These backgrounds make up just over 1\% of the $\fL$ signal yield, and their contributions are estimated using
control samples in data, as described in Section~\ref{sec:NonpromptProcedure}.

The nonprompt background that we estimate from data includes only the contributions coming from $\ZtoLL$ and
$\ttbar \to 2\Pell 2\PGn$ events.
Other processes such as diboson production also contribute events with one or more nonprompt leptons,
but their impact is much smaller and better modeled by simulation than the $\ZtoLL$ and $\ttbar$ contributions.
Therefore, the diboson backgrounds are estimated together with the prompt processes that we estimate from simulation.
These prompt backgrounds include triboson, Higgs boson, and $\ttbar\PZ$ production,
and altogether they comprise about 1\% of the $\fL$ yield.
Backgrounds from leptonic $\tau$ decays, estimated from MC simulations, are found to be negligible.

For the $\ZtoLL$ process, the total background contribution is below 0.5\% in both the $\MM$ and $\EE$
final states. We estimate all $\ZtoLL$ backgrounds from simulation, including diboson, $\ttbar(+\PZ)$,
triboson, $\PZ\to\tau^+\tau^-$, and Higgs boson production.

\subsection{Nonprompt-lepton background estimate}  \label{sec:NonpromptProcedure}

Events with nonprompt leptons will not necessarily lead to the combination of charges in an $\LL\LL$ final state.
Therefore, we use a same-sign method to estimate the number of selected events with nonprompt leptons.
A control region of events with one same-flavor, opposite-sign lepton pair $\Zo$~(consisting of prompt leptons)
and one same-flavor, same-sign lepton pair $\Zt$~(consisting of nonprompt leptons) is defined. With the exception
of the same-sign requirement for $\Zt$, the selection requirements for this control region are identical to those
described in Section~\ref{sec:Selection} for the $\fL$ signal region. In order to match the signal region as closely
as possible, the requirement $m_{\Zt} > 4\GeV$ is imposed.

We estimate the nonprompt background, $\Nnpr$, as $ \Nnpr = \lambda(\Nobs - \Npr) $
where $\lambda$ scales the number of nonprompt events from the control region to the signal region,
$\Nobs$ is the number of observed data events in the control region, and $\Npr$ is the total number
of expected events from the simulated prompt processes in the control region. We take $\lambda = 1$;
that is, we assume all charge configurations of nonprompt leptons are equally probable.
This assumption is supported by studies of simulated $\ZtoLL$ and $\ttbar \to 2\Pell 2\PGn$ events,
and we assign a systematic uncertainty of 30\% to $\lambda$, based on these studies.
The statistical uncertainty of $\Nnpr$ is assessed by propagating the statistical uncertainties of
$\Npr$ and $\Nobs$.

When measuring the differential decay rates, it is necessary to characterize the shapes of the nonprompt
contributions in order to subtract them and this requires a sufficiently large sample of background events.
For this purpose, an additional ``loose'' control region is defined. This region is identical to the
one described above---which, for clarity, we will refer to as the ``tight'' control region---except that
we also allow events with one lepton that passes loosened identification criteria. This lepton must pass
the ``loose'' criteria listed in Section~\ref{sec:Objects} and it must be in the $\Zt$ pair. The other
three leptons in the event are required to pass the tight identification criteria. Events with more than
one lepton candidate passing the loose but failing the tight identification criteria are rejected.

We apply an overall normalization to the loose control region to correct for the different selection
efficiencies of loosely and tightly identified leptons. The yield of the loose control region is scaled
by the factor $ \lambda\loose = {\Nnpr} / {\Nnpr\looseCR} $,
where $\Nnpr$ and $\Nnpr\looseCR$ are the numbers of nonprompt events in the tight and loose control regions.

\section{Systematic uncertainties} \label{sec:Systematics}
Many sources of systematic uncertainty impact the $\ZtofL$ and $\ZtoLL$ yields similarly, and are therefore nearly
or completely canceled in the branching fraction measurement given by Eq.~\refeq{eqn:BranchingFraction}.
These sources of uncertainty concern $\PZ$ production and include, for example, the integrated luminosity measurements
and pileup modeling. The most significant sources of systematic uncertainty are therefore those that
depend on the number of selected leptons; namely, the lepton identification and reconstruction efficiencies.
The assessed systematic uncertainties in the $\BFfM$, $\BFtMtE$, and $\BFfE$ measurements are listed in
Table~\ref{tab:Systematics}.

\begin{table}[ht] \centering
\topcaption{Systematic uncertainties in the branching fraction, listed by source and by final state.
The uncertainty from each source varies by data-taking period. Uncertainties from pileup and theory
sources are identical for all final states.}
\begin{scotch}{l l l l l }
        & \multicolumn{4}{c}{Uncertainty in $\BF$~(\%)} \\
        Source && $\fM$ \hspace{1cm} & $\tMtE$ \hspace{1cm} & $\fE$ \hspace{1cm} \\
\hline
        Muon identification && 2 & 1 & 0 \\
        Electron identification && 0 & 1--2 & 3--4 \\
        Electron reconstruction && 0 & 1--3 & 2--6 \\
        Background (statistical) && $<$1 & 1--6 & 2--6 \\
        Background (systematic) && $<$1 & 1--5 & 1 \\
        Trigger efficiency && 0.2--2 & 1--1.5 & 1--2 \\[\cmsTabSkip]
        Pileup distribution && 0.5--1 & 0.5--1 & 0.5--1 \\
        QCD $\muR$ and $\muF$ scales && 0.8 & 0.8 & 0.8 \\
        PDF sets && 0.6 & 0.6 & 0.6 \\
\end{scotch}
\label{tab:Systematics}
\end{table}

The efficiency to identify or reconstruct a lepton of a particular flavor, in accordance with the requirements
listed in Section~\ref{sec:Objects}, is measured using the ``tag-and-probe'' procedure~\cite{TagAndProbe}.
These efficiencies are measured for finite regions of the fiducial detector volume, defined by $\ptL$ and $\etaL$ of the relevant leptons.
We assess the uncertainties from muon identification, electron identification, and electron reconstruction by generating several hundred sets of alternate data-to-simulation scale factors for each uncertainty source and each data-taking period. Each alternate scale factor is normally distributed about its central value with standard deviation equal to its statistical uncertainty. We also include a fully-correlated 1\% uncertainty among the efficiency measurements in each set. For all data-taking periods, the uncertainty from muon identification is about 2\% in the $\fM$ final state. In the $\fE$ final state, the electron identification uncertainty varies from about 3 to 4\%, and the reconstruction uncertainty ranges 2--6\%. The $\tMtE$ final state has contributions of 1--3\% from each of the three sources.

We assess a 30\% systematic uncertainty to the nonprompt-lepton background estimate from control samples
in data described in Section~\ref{sec:NonpromptProcedure}.
This assessment is based on observed variations of the estimate as we vary the details of the procedure for
calculating it. These variations are caused by the way sources of nonprompt leptons contribute
to the control sample. This systematic uncertainty is assigned in addition to the statistical uncertainty in
the total number of estimated nonprompt lepton background events.
Both contributions are listed in Table~\ref{tab:Systematics} and range from $<1$ to 6\% for the
various $\fL$ final states. The uncertainty from the efficiency of each type of dilepton trigger is assessed
as half of the difference of the data-to-simulation efficiency ratio from unity.
The efficiencies of the dimuon and dielectron triggers used in the analysis are not identical in data and in simulation,
although their ratio is close to unity in all cases. For events of the $\tMtE$ final state, which may
satisfy either the dimuon or the dilepton trigger, we use the average of the two differences.
The overall uncertainty from trigger efficiency is 1--2\%.
Uncertainties from the pileup conditions are evaluated by varying the cross section controlling the level of pileup
in the simulation up or down by one standard deviation. Correlations from one data-taking period to another are taken
into account. The corresponding uncertainty in $\BF$ is below 1\% for all data-taking periods and is identical for
all $\fL$ final states.

Theoretical uncertainties can impact the measurement of $\BF$ through the calculation of the
ratio of acceptances appearing in Eq.~(\ref{eqn:BranchingFraction}).
These uncertainties arise from variations in the renormalization $\muR$ and
factorization $\muF$ scales and from imprecise knowledge of the PDFs.
The uncertainty from these sources is estimated by reweighting the simulated $\ZtofL$ and $\ZtoLL$
event samples. The alternate weights for the renormalization and factorization scales are
calculated by varying $\muR$ and $\muF$ independently by factors of $1/2$ and by 2, excluding the cases
where one is multiplied by $1/2$ and the other by 2. One-half the difference between the largest and
smallest values is taken to be the systematic uncertainty and it amounts to less than 1\%.
The alternate PDF sets are 100 variations of \NNPDF~3.1 (or \NNPDF~3.0 for some simulated data sets), corresponding
to an uncertainty in the branching fraction of 0.6\%. The systematic uncertainties from
both theory sources are identical for all $\fL$ final states.

\section{Results} \label{sec:Results}
The results of this analysis begin with the event yields and the $\ZtofL$ branching fraction measurements.
Next, nine differential $\ZtofL$ decay rate measurements are reported, followed by the results of probes for \CP violation
in the $\ZtofL$ decay process. Finally, limits on new physics in the $\ZtofL$ channel are presented.

Numerical values for the results reported in this paper are provided in HEPData~\cite{HEPData}.

\begin{figure*}[htp] \centering
\includegraphics[width=\cmsFigWidth]{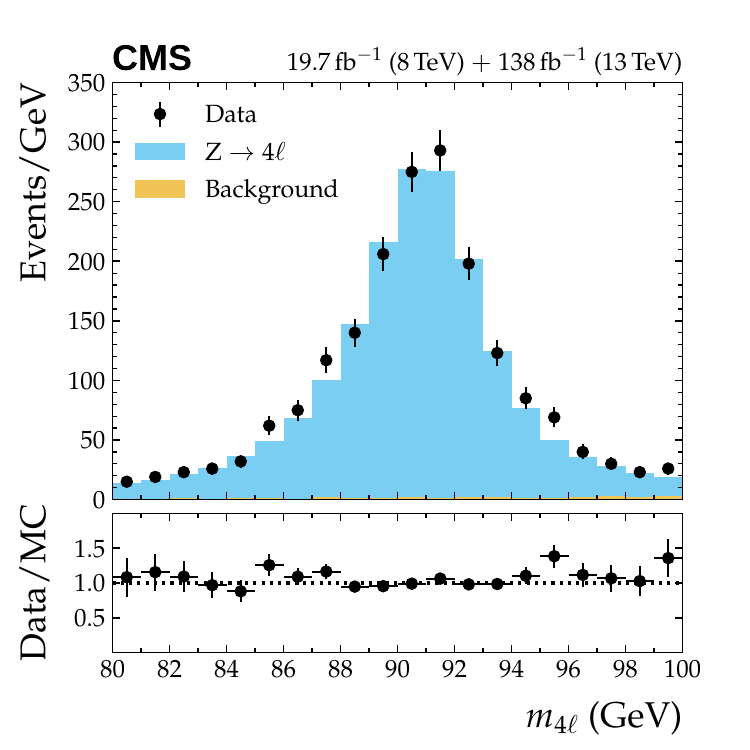}
\includegraphics[width=\cmsFigWidth]{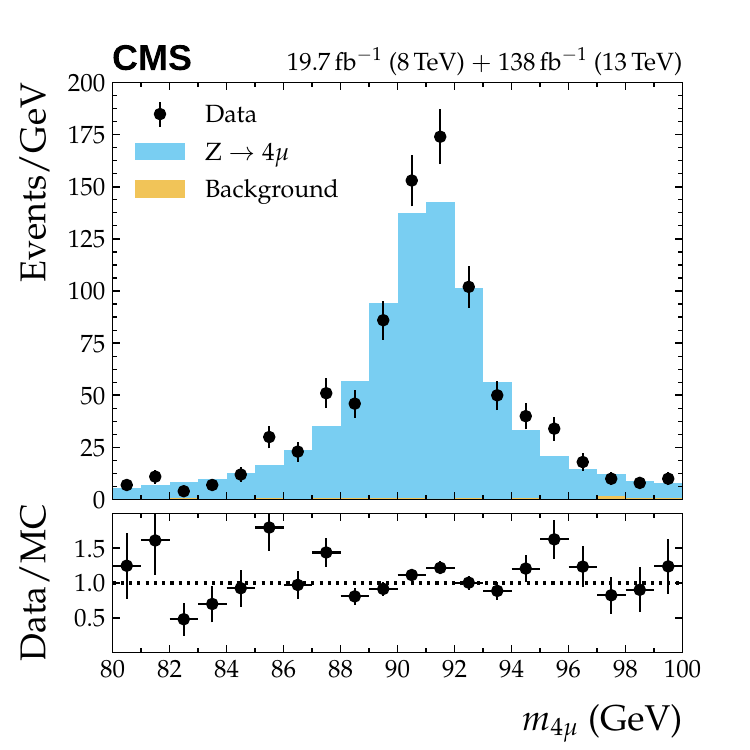} \\
\includegraphics[width=\cmsFigWidth]{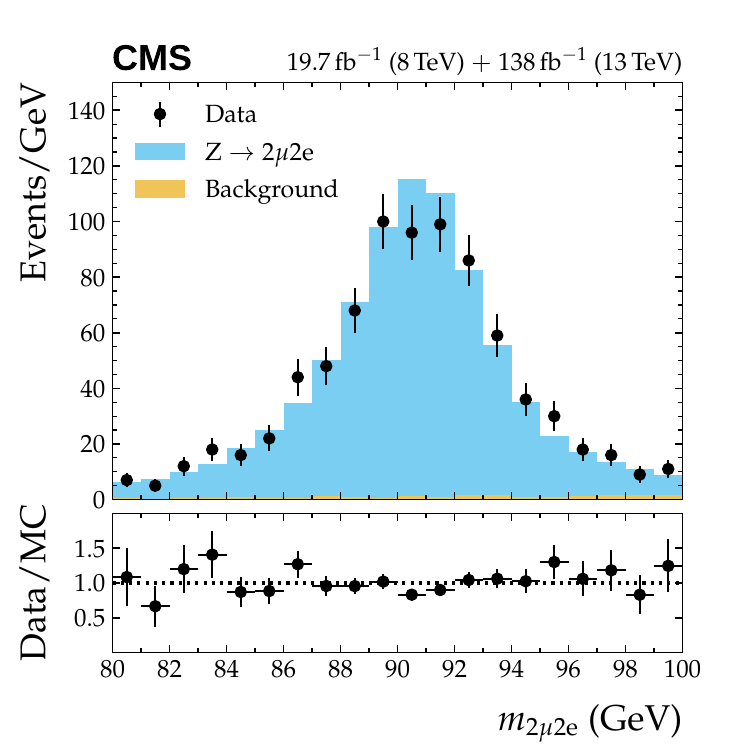}
\includegraphics[width=\cmsFigWidth]{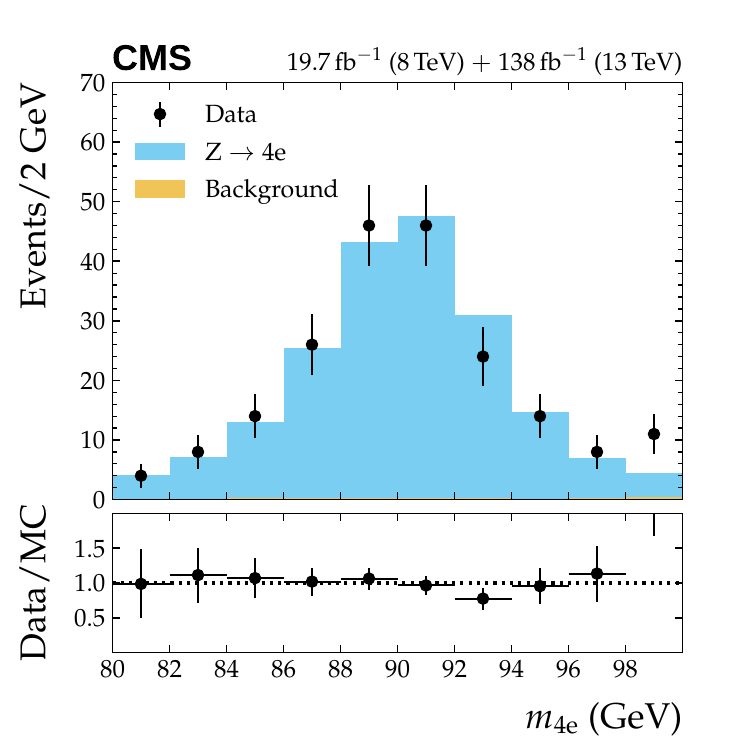} \\
\caption{Invariant mass distributions for events passing the $\ZtofL$ selection for
the sum of all $\fL$ channels~(upper left), the $\fM$ channel~(upper right), the $\tMtE$~channel (lower left),
and the $\fE$ channel~(lower right).
Vertical error bars represent the total uncertainty on each point.
The simulations are normalized according to predicted cross sections and the integrated luminosity. }
\label{fig:Mass4L}
\end{figure*}

The expected sample composition after applying the $\ZtofL$ selection described in Section~\ref{sec:Selection}
is given in Table~\ref{tab:Yield4L}. The sample composition of $\ZtoLL$ events is listed in Table~\ref{tab:YieldLL}.
Distributions of the invariant mass $\mfL$ for all $\ZtofL$ events are shown in Fig.~\ref{fig:Mass4L}.

\begin{table*} \centering
\topcaption{Observed event yields and sample composition for the $\ZtofL$ signal region
listed for the sum of all $\fL$ channels and for each channel individually.
These yields are inclusive of all 8 and 13\TeV data samples.
The uncertainties shown are obtained by combining statistical and systematic uncertainties in quadrature.}
\begin{scotch}{l l l r@{ $\pm$ }r r@{ $\pm$ }r r@{ $\pm$ }r r@{ $\pm$ }r}
        \multicolumn{3}{l}{Type} & \multicolumn{2}{l}{$N_{\fM}$} & \multicolumn{2}{l}{$N_{\tMtE}$} & \multicolumn{2}{l}{$N_{\fE}$} & \multicolumn{2}{l}{$N_{\fL}$ (total)} \\
\hline
        \multicolumn{3}{l}{Observed} & \multicolumn{2}{l}{876} & \multicolumn{2}{l}{800} & \multicolumn{2}{l}{201} & \multicolumn{2}{l}{1\,877} \\
        \multicolumn{3}{l}{Expected} & 806.4 & 44.0 & 806.0 & 43.2 & 198.1 & 10.9 & 1\,810.5 & 62.6 \\
        &       \multicolumn{2}{l}{Signal} & 793.4 & 43.9 & 784.6 & 42.9 & 195.1 & 10.7 & 1\,773.1 & 62.3 \\
        &       \multicolumn{2}{l}{Background} & 12.9 & 3.4 & 21.4 & 4.9 & 3.0 & 1.9 & 37.4 & 6.3 \\
        &       &       Nonprompt & 8.4 & 3.4 & 16.2 & 4.9 & 1.3 & 1.9 & 25.9 & 6.3 \\
        &       &       Other & 4.5 & 0.4 & 5.2 & 0.3 & 1.8 & 0.3 & 11.5 & 0.6 \\[\cmsTabSkip]
        \multicolumn{3}{l}{Signal purity (\%)} & 98.4 & 1.2 & 97.3 & 1.1 & 98.5 & 2.3 & 97.9 & 0.6 \\
        \multicolumn{3}{l}{$N^\text{obs} / N^\text{exp}$ (\%)} & 108.6 & 3.8 & 99.3 & 3.6 & 101.4 & 7.4 & 103.7 & 2.4 \\
\end{scotch}
\label{tab:Yield4L}
\end{table*}

\begin{table*} \centering
\topcaption{Observed event yields and sample composition for the $\ZtoLL$ signal regions,
inclusive of all 8 and 13\TeV data samples.
The uncertainties shown are obtained by combining statistical and systematic uncertainties in quadrature.}
\begin{scotch}{l l l r@{ $\pm$ }r r@{ $\pm$ }r r@{ $\pm$ }r}
	\multicolumn{3}{l}{Type} & \multicolumn{2}{l}{$N_{\MM}$} & \multicolumn{2}{l}{$N_{\EE}$} & \multicolumn{2}{l}{$N_{\LL}$ (total)} \\
        \hline
        \multicolumn{3}{l}{Observed} & \multicolumn{2}{l}{105,025,896} & \multicolumn{2}{l}{75,569,408} & \multicolumn{2}{l}{180,595,340} \\
        \multicolumn{3}{l}{Expected} & 105,300,000 & 2,900,000 & 77,220,000 & 2,100,000 & 182,500,000 & 3,600,000 \\
	&	\multicolumn{2}{l}{Signal} & 104,900,000 & 2,900,000 & 76,940,000 & 2,100,000 & 181,900,000 & 3,600,000 \\
	&	\multicolumn{2}{l}{Background} & 345,000 & 9,900 & 273,000 & 7,900 & 618,000 & 13,000 \\
	&	&	Diboson & 193,500 & 5,500 & 150,900 & 4,300 & 344,500 & 6,900 \\
	&	&	$\ttbar$ & 143,800 & 8,200 & 115,900 & 6,600 & 259,700 & 11,000 \\
	&	&	Other & 7,682 & 280 & 6,126 & 230 & 13,810 & 360 \\[\cmsTabSkip]
	\multicolumn{3}{l}{Signal purity (\%)} & 99.672 & 0.002 & 99.646 & 0.003 & 99.661 & 0.001 \\
	\multicolumn{3}{l}{$N^\text{obs} / N^\text{exp}$ (\%)} & 99.759 & 0.010 & 97.866 & 0.011 & 98.958 & 0.007 \\
\end{scotch}
\label{tab:YieldLL}
\end{table*}

\subsection{Branching fraction measurements} \label{sec:Measurements}

The individual branching fractions $\BFfM$, $\BFtMtE$, and $\BFfE$ are measured using each of the 2012, 2016, 2017,
and 2018 data samples. The 8\TeV data, taken in 2012, are expected to contribute 8.4\% of the total yield.
Simultaneous fits to these 12 measurements are performed with a procedure that is
equivalent to the best linear unbiased estimation method for combining measurements~\cite{BLUE}.
The free parameters in the fit are signal strength parameters that scale the SM expected value of
the branching fraction to the observed one. The measured value is the product of the signal strength
and the expected value.
Two fits are performed: one in which a single signal strength is used to parameterize all
$\fL$ final states simultaneously (the ``SM'' parameterization), and another
in which the signal strengths for the $\fM$, $\tMtE$, and $\fE$
final states are allowed to vary independently~(the ``BSM'' parameterization).
The SM parameterization assumes lepton flavor universality in the $\ZtofL$ process and is the nominal result.
The BSM parameterization is used for the new physics constraints reported in Section~\ref{sec:Limits}.

The fitting procedure includes correlations of systematic uncertainties among the measurements.
We consider each source to be completely correlated for all measurements within a particular data-taking period.
Lepton identification and reconstruction efficiencies are fully correlated among the 13\TeV data and uncorrelated with the 8\TeV data,
with the exception of a 50\% correlation for electron identification efficiency between 8 and 13\TeV.
In general, the dimuon and dielectron trigger efficiencies are uncorrelated with each other,
and all trigger efficiencies are uncorrelated among the data-taking periods.
The systematic uncertainty from background estimation is fully correlated across data-taking periods,
as are all uncertainties from theory sources.

The fit results are listed in Table~\ref{tab:BSMMeasurements}. The values for the individual channels
$\fM$, $\tMtE$, and $\fE$ are statistically independent and are based on the BSM parameterization.
The value for the combined channel $\fL$ is not simply the sum of the values for the individual channels
because it comes from the SM fit which is based on a parameterization different from that of the BSM fit.
This measurement of $\BF$ has a precision of about 3.2\% which is substantially more precise than the
combined value reported in Ref.~\cite{Rainbolt}
and the latest value reported by the Particle Data Group~\cite{PDG24}.
It is twice as precise as the most recent measurement reported by the ATLAS Collaboration~\cite{ATLAS21}.
All measurements in Tables~\ref{tab:BSMMeasurements} are consistent with the SM predictions.

\begin{table} \centering
\topcaption{Measured branching fractions.
The measured values for individual channels $\fM$, $\tMtE$, and $\fE$ are obtained from the BSM fit
described in Section~\ref{sec:Measurements}. The value for the combined channel $\fL$ is obtained
from the SM fit.}
\begin{scotch}{c l@{ $\pm$ }l l@{ $\pm$ }l@{ \stat} @{ $\pm$ }l@{ \syst}}
	& \multicolumn{5}{c}{$\BF$ [$\ten{$-6$}$]} \\
	Channel & \multicolumn{2}{c}{Expected} & \multicolumn{3}{c}{Observed} \\
\hline
	$\fM$   & 1.20 & 0.01 & 1.25 & 0.04 & 0.03 \\
	$\tMtE$ & 2.31 & 0.02 & 2.17 & 0.08 & 0.06 \\
	$\fE$   & 1.20 & 0.01 & 1.16 & 0.09 & 0.06 \\ [\cmsTabSkip]
	$\fL$   & 4.70 & 0.05 & 4.67 & 0.11 & 0.10 \\
\end{scotch}
\label{tab:BSMMeasurements}
\end{table}

\subsection{Differential decay rates} \label{sec:DifferentialDistributions}

We report the differential decay rates $\dGdx$, where $x$ is one of nine kinematic and angular quantities.
These quantities are, unless otherwise specified, measured in the rest frame of the $\PZ$ boson, which is taken to be
the center-of-mass frame of the reconstructed four-lepton system.

For each observed data distribution, we subtract the expected background bin-by-bin.
The shape of the nonprompt-lepton background is taken from the loose control region described in Section~\ref{sec:NonpromptProcedure}
and the shapes of other backgrounds are taken from simulations. After background subtraction,
each distribution is unfolded to the dressed lepton level in order to correct for the effects of detector resolution.
The software package {\textsc{PyUnfold}}~\cite{PyUnfold} is used, which implements the D'Agostini method of iterative
unfolding~\cite{DAgostini} with a customizable initial prior distribution. We begin with a flat prior and
iterate until the result converges. Convergence is achieved when the covariance matrix remains stable between iterations
within a 1\% tolerance, which is met before 50 iterations for each distribution. For the quantities considered here,
resolution effects are small and unfolding has a minor effect on the distributions.
The unfolded distributions are corrected to the full phase space by taking into account the acceptance and efficiency
for each bin in the distribution. Each distribution is normalized such that its integral is equal to
\begin{equation*}
    \PWfL= \BF \FWZ = 12.83 \pm 0.41 \keV,
\end{equation*}
where $\BF$ is the fit result given in Table~\ref{tab:BSMMeasurements} and $\FWZ = 2.4955 \pm 0.0023\GeV$ is
the world-average total \PZ\,boson width~\cite{PDG24}.

\begin{figure*} \centering
\includegraphics[width=\cmsFigWidth]{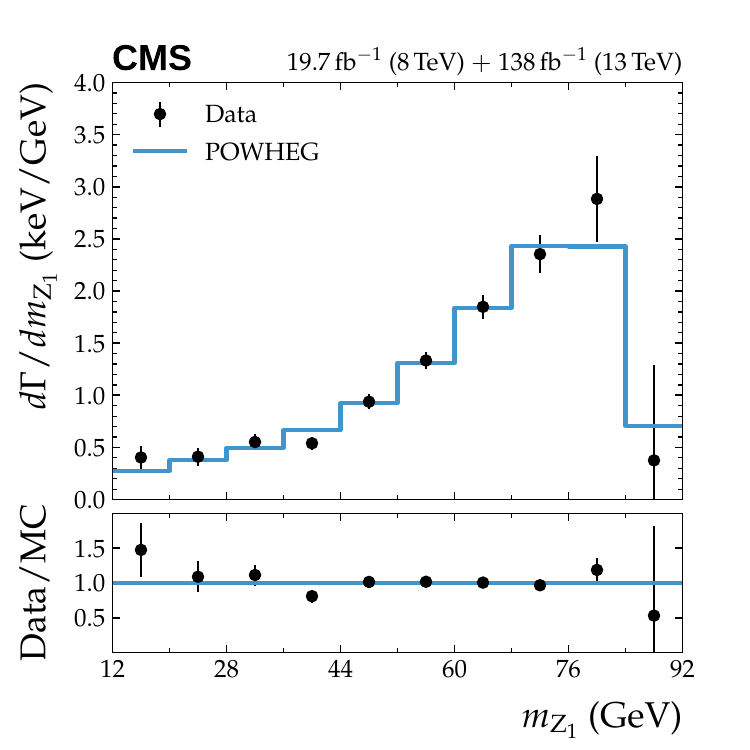} 
\includegraphics[width=\cmsFigWidth]{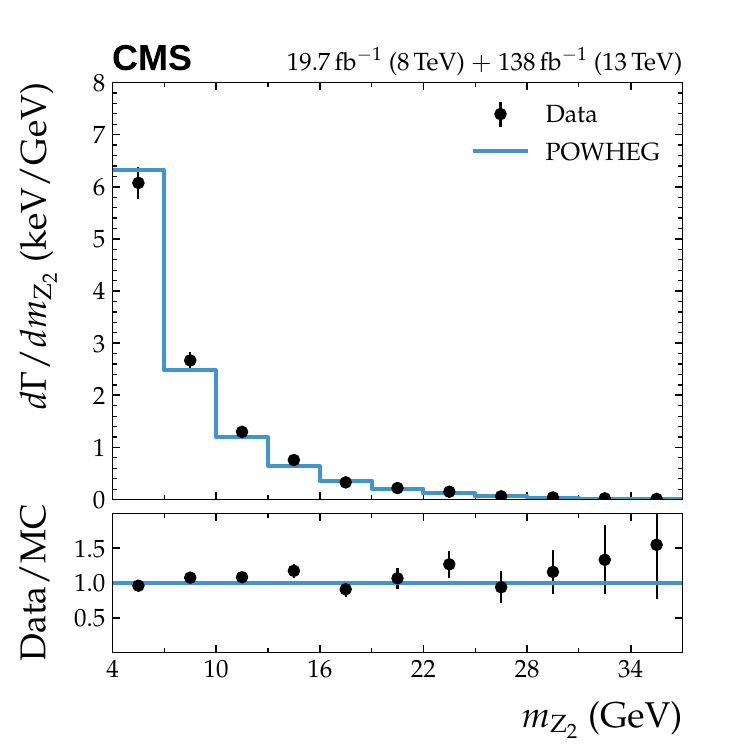} \\
\includegraphics[width=\cmsFigWidth]{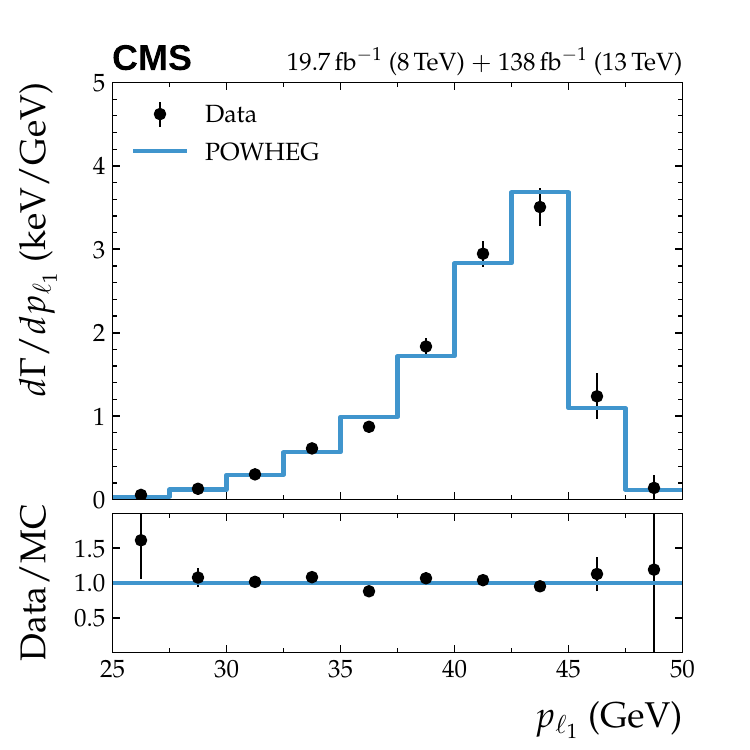} 
\includegraphics[width=\cmsFigWidth]{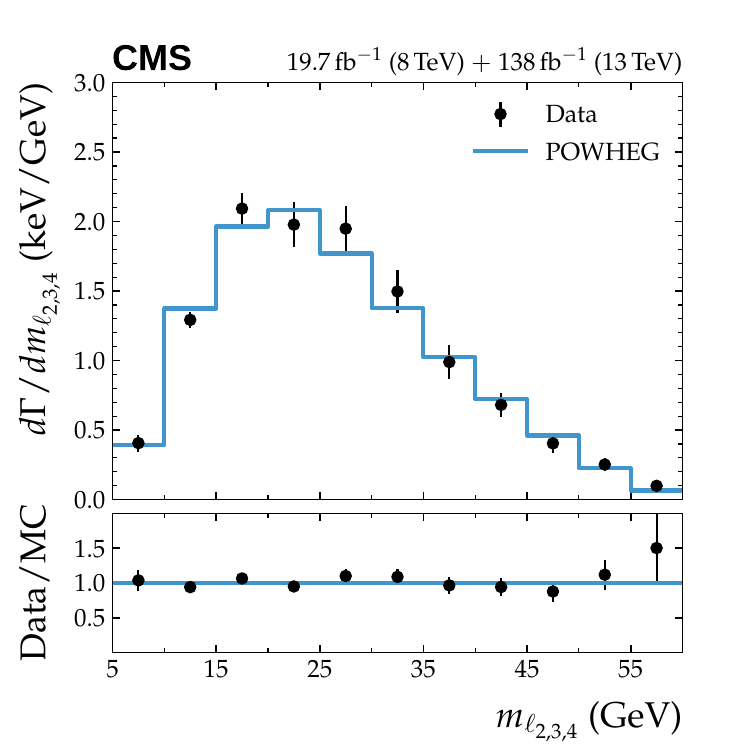} \\
\caption{The differential $\ZtofL$ decay rate $\dGdx$ as a function of the quantities $\mZo$~(upper left),
$\mZt$~(upper right), $\pLo$~(lower left), and $\mtrlttt$~(lower right) defined in Section~\ref{sec:DifferentialDistributions}.
These distributions have been unfolded and scaled to the
full phase space region using the $\BF$ result of Table~\ref{tab:BSMMeasurements}.
For each bin, the total uncertainty is shown.}
\label{fig:KinematicDDR}
\end{figure*}

The differential decay rates for the invariant masses of the lepton pairs, $\dGd{\mZo}$ and $\dGd{\mZt}$, are shown in
Fig.~\ref{fig:KinematicDDR}, upper left and upper right, respectively. The phase space
definition requires $\mZt > 4\GeV$, and the event selection requires $\mZo > 12\GeV$.
Figure~\ref{fig:KinematicDDR}~(lower left) shows $\dGd{\pLo}$, where $\pLo$ is the momentum of the leading lepton in the
\PZ~boson rest frame. The distribution of $\pLo$ peaks sharply just below 45\GeV, which is approximately half the
nominal mass of the \PZ~boson. Figure~\ref{fig:KinematicDDR}~(lower right) shows $\dGd{\mtrlttt}$, where $\mtrlttt$ is the
invariant mass of the trailing three leptons; this quantity tends to be anticorrelated with $\pLo$.
Its distribution falls sharply below 20\GeV due to the $\mLL > 4\GeV$ phase space definition.

\begin{figure*} \centering
\includegraphics[width=0.9\cmsFigWidth]{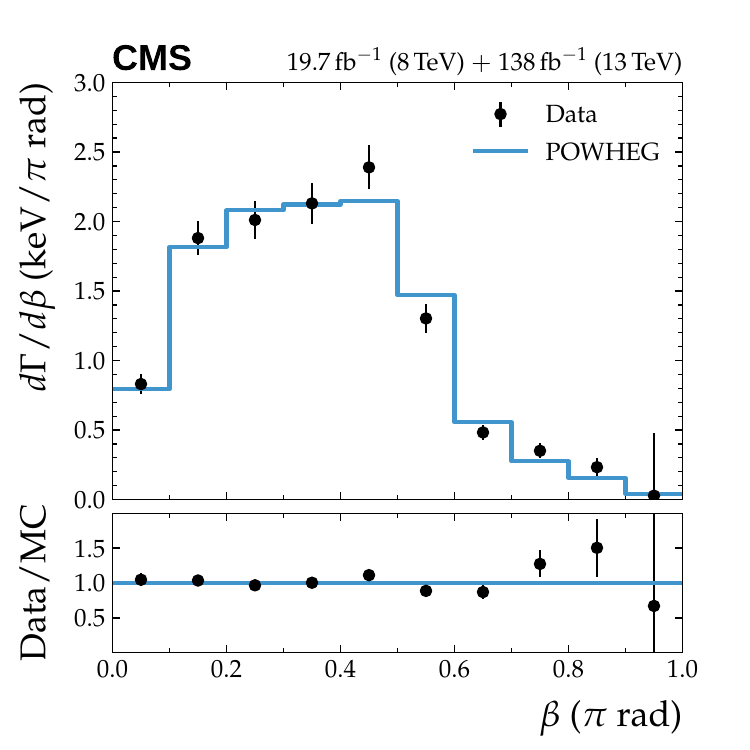} \\
\includegraphics[width=0.9\cmsFigWidth]{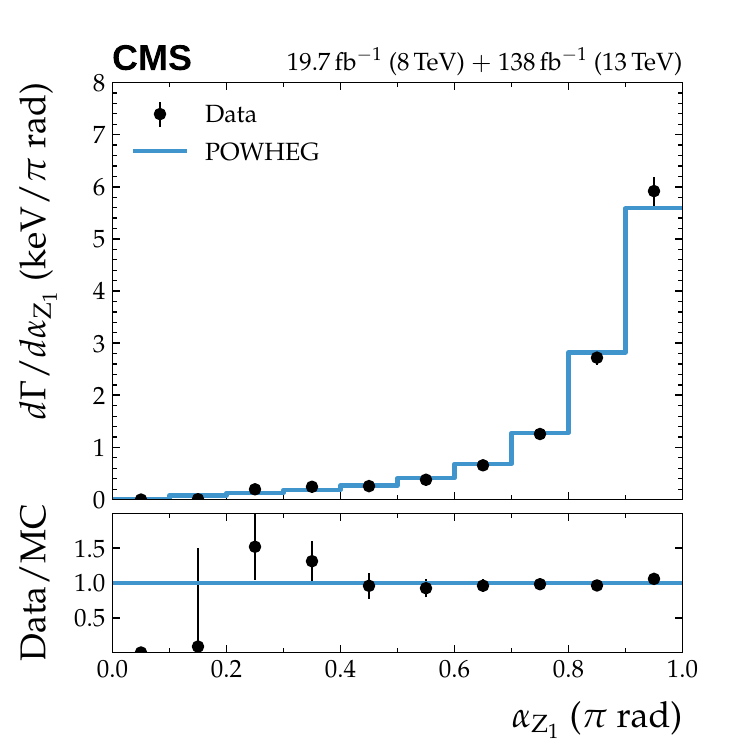} 
\includegraphics[width=0.9\cmsFigWidth]{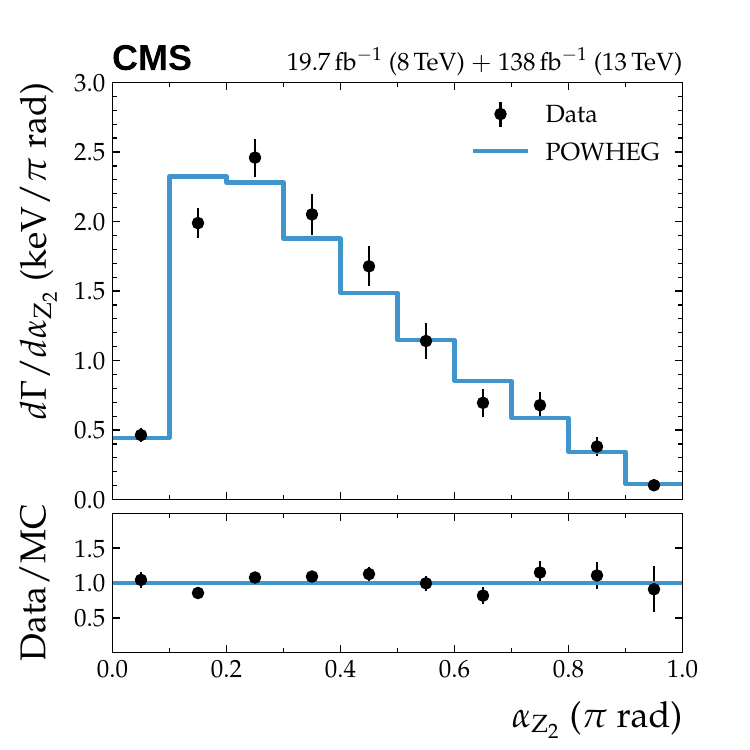} \\
\includegraphics[width=0.9\cmsFigWidth]{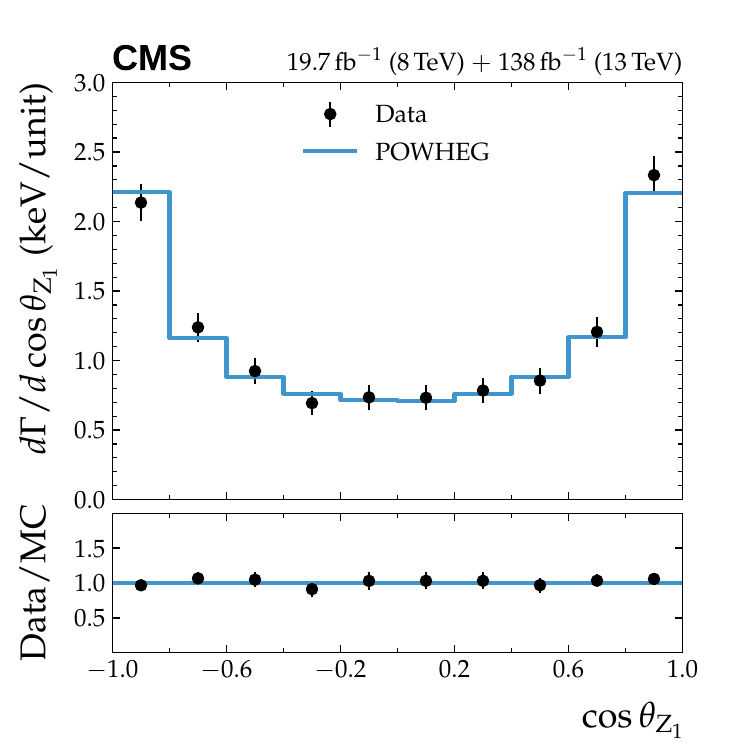} 
\includegraphics[width=0.9\cmsFigWidth]{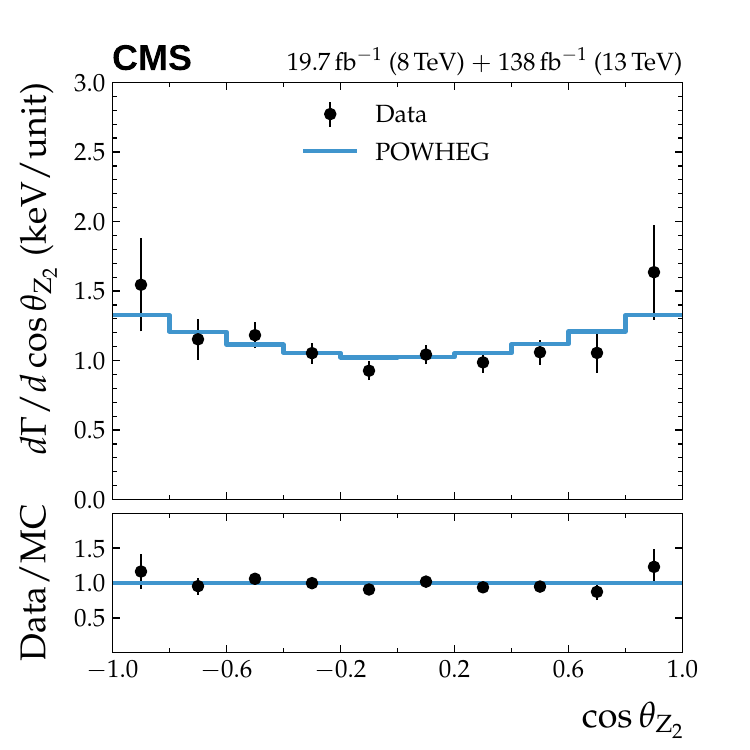}
\caption{The differential decay rates $\dGd{\beta}$~(upper),
$\dGd{\alpha_{\Zo}}$~(middle left), $\dGd{\alpha_{\Zt}}$~(middle right),
$\dGd{\cos\theta_{\Zo}}$~(lower left), and $\dGd{\cos\theta_{\Zt}}$~(lower right).
Here, $\beta$ and $\alpha_{\Zo}$ have been scaled so that $\pi$ radians
corresponds to 1.0 on the plot.
These distributions have been unfolded and scaled to the
full phase space region using the $\BF$ result of Table~\ref{tab:BSMMeasurements}.
For each bin, the total uncertainty is shown.}
\label{fig:AngularDDR}
\end{figure*}

The measurement of $\dGd{\beta}$ is shown in Fig.~\ref{fig:AngularDDR}~(upper), where $\beta$ is the angle between
the momentum of the less-energetic lepton of the $\Zo$ pair and the momentum of $\Zt$. This angle quantifies the
collimation of the virtual particle emission shown in Fig.~\ref{fig:FeynmanDiagram}.
Distributions of $\dGd{\alpha_{\Zo}}$ and $\dGd{\alpha_{\Zt}}$ are shown in Fig.~\ref{fig:AngularDDR},
middle left and middle right, respectively, where $\alpha_{\Zo}$~($\alpha_{\Zt}$) is the angle between the momenta
of the leptons of pair $\Zo$~($\Zt$) in the \PZ rest frame. The leptons of $\Zo$ are expected to be well separated,
which is confirmed in Fig.~\ref{fig:AngularDDR}~(middle left). However, they cannot be fully back-to-back because of
the presence of $\Zt$, and the distribution of $\alpha_{\Zo}$ in fact extends below $\pi$. The $\Zt$ leptons, by contrast,
are expected to be collimated, but their angular separation cannot be arbitrarily small because of the requirement that $m_{\Zt} > 4\GeV$.
Distributions of $\dGd{\cos\theta_{\Zo}}$ and $\dGd{\cos\theta_{\Zt}}$ are shown in Fig.~\ref{fig:AngularDDR}, lower left
and lower right, respectively. Here $\theta_{\Zo}$~($\theta_{\Zt}$) is the angle between the momentum of the positively-charged
lepton of pair $\Zo$~($\Zt$) and the momentum of $\Zt$~($\Zo$), measured in the rest frame of $\Zo$~($\Zt$).
These angles characterize the helicity of each lepton pair, and the observed symmetry of the distributions of
$\cos\theta_{\Zo}$ and $\cos\theta_{\Zt}$ reflects the \CP invariance of the $\ZtofL$ decay process.

Overall, the agreement of the simulation with the measured differential decay rates
displayed in Figs.~\ref{fig:KinematicDDR} and~\ref{fig:AngularDDR} is good.

\subsection{Measurement of the triple-product asymmetry}

The SM does not predict observable violations of \CP invariance in decays of \PZ bosons,
so any evidence for \CP violation would constitute a clear signal of BSM physics.
Generally, \CP violation in decays of unstable particles can
appear as a difference between the amplitudes for the decay to a specific final state
and to its \CP-conjugate state~\cite{Bensalem,Bevan}. Assuming that all gauge interactions
respect {$\mathit{CPT}$} invariance, where {$\mathit{T}$} is time-reversal symmetry,
a T-odd observable such as the scalar triple product
$X = (\vec{p}_a \times \vec{p}_b) \cdot \vec{p}_c$ can be used to test for \CP invariance
violation that would be manifested as an asymmetry in $X$~\cite{Bensalem}.
In the present case of four-body \PZ boson decays, decomposed as two-body decays
to $\Zo$ and $\Zt$, the angle $\phi$ between the respective decay planes provides
an opportunity to test for \CP invariance violation by measuring the asymmetry
$\Asinphi$ defined in Section~\ref{sec:CPViolationProcedure}.

\begin{figure*}[hb]
\centering
\includegraphics[width=\cmsFigWidth]{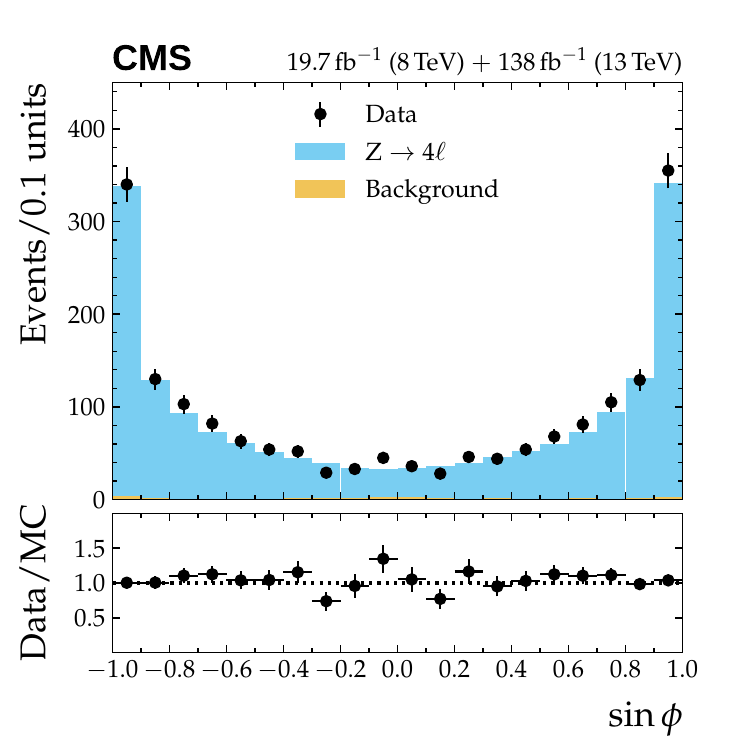}
\includegraphics[width=\cmsFigWidth]{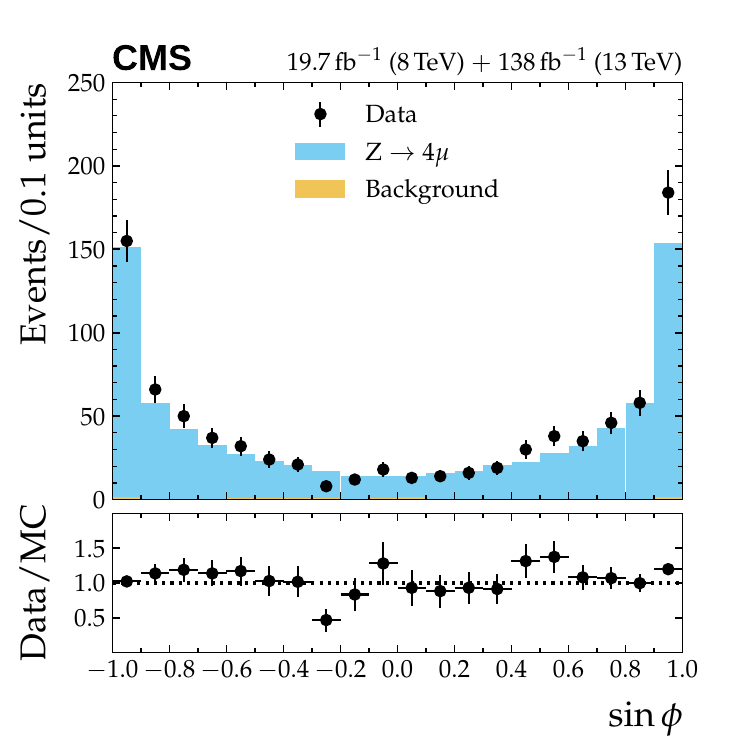} \\
\includegraphics[width=\cmsFigWidth]{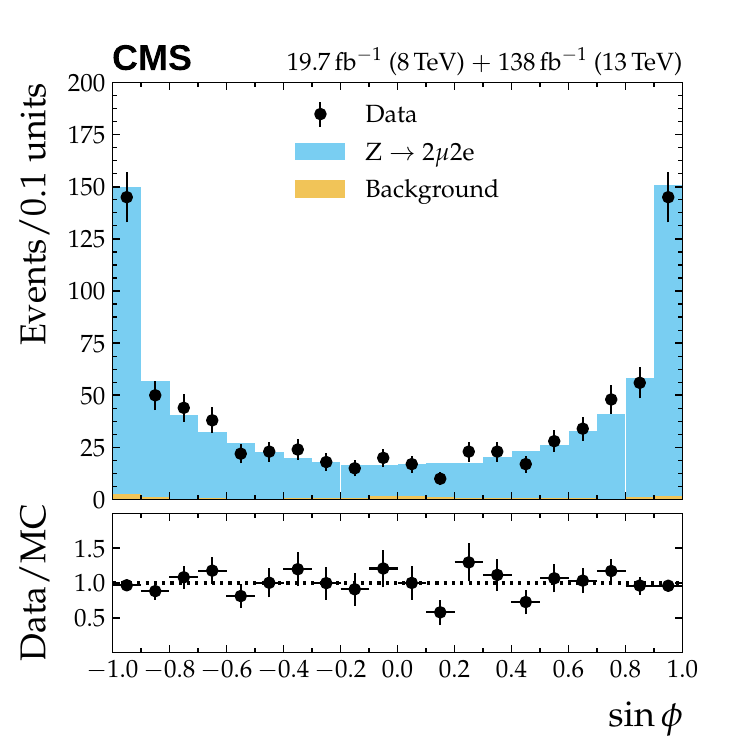}
\includegraphics[width=\cmsFigWidth]{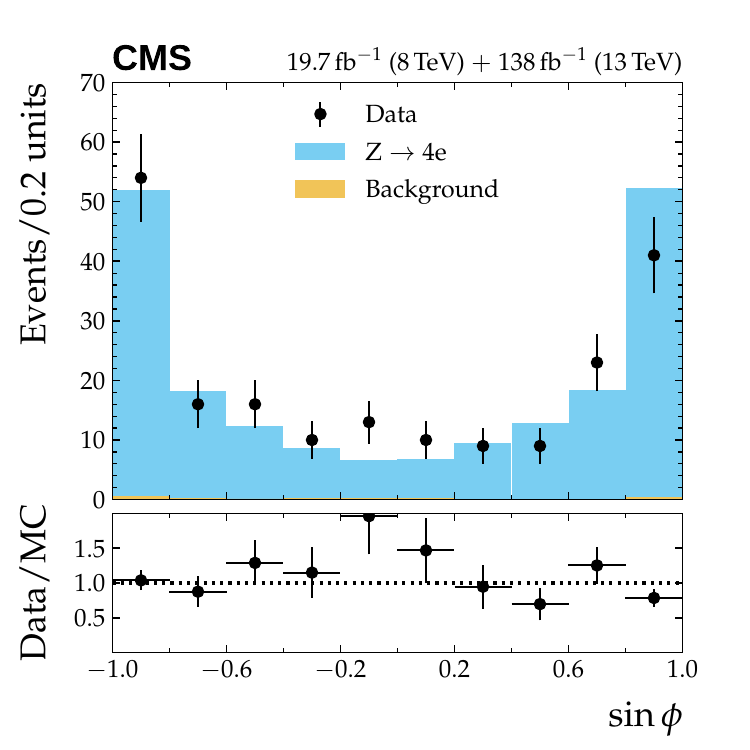} \\
\caption{Distributions of $\sin\phi$ for the sum of all $\fL$ channels~(upper left), the $\fM$ channel~(upper right),
the $\tMtE$ channel~(lower left), and the $\fE$ channel~(lower right).
Vertical error bars represent the total uncertainty on each point.
The simulations are normalized according to predicted cross sections and the integrated luminosity.}
\label{fig:SinPhi}
\end{figure*}

The measurements of $\Asinphi$ for all $\fL$ final states are given in Table~\ref{tab:Asymmetry}. Every measured value
is consistent with the SM expectation of zero, hence there is no evidence of \CP invariance violation.
The table also lists $N_+$ and $N_-$, the numbers of events with
$\sin\phi > 0$ or $< 0$, respectively, after subtracting background contributions
determined from the loose control region described in Section~\ref{sec:Background}. The background estimation is the only
nonnegligible source of systematic uncertainty for $\Asinphi$, which is much smaller than the statistical uncertainty.
Both the statistical and systematic uncertainties for $\Asinphi$ are computed taking $\Asinphi = 0$.
The distributions of $\sin\phi$ are shown in Fig.~\ref{fig:SinPhi} for all $\ZtofL$ events and for each final state studied.

\begin{table}[ht] \centering
\topcaption{The number of signal $\ZtofL$ events $N_+$~($N_-$) in which $\sinphi > 0$\,($< 0$),
and the associated asymmetry $A_{\sinphi}$ defined in Eq.~\refeq{eqn:TripleProductAsymmetry}.
The yields $N_+$ and $N_-$ are reported after subtraction of backgrounds. Uncertainties are
computed with respect to $A_{\sinphi} = 0$.}
\begin{scotch}{c r r l@{ $\pm$ }l@{ \stat} @{ $\pm$ }l@{ \syst}}
Channel & \multicolumn{1}{c}{$N_{+}$} & \multicolumn{1}{c}{$N_{-}$} & \multicolumn{3}{c}{$A_{\sin\phi}$ (\%)} \\
\hline
$4\PGm$ & 412 & 450 & $-4.3$ & 3.4 & 0.3 \\
$2\PGm 2\Pe$ & 372 & 406 & $-4.3$ & 3.6 & 0.4 \\
$4\Pe$ & 98 & 101 & $-1.4$ & 7.1 & 0.6 \\ [\cmsTabSkip]
$4\Pell$ & 882 & 956 & $-4.0$ & 2.3 & 0.2 \\
\end{scotch}
\label{tab:Asymmetry}
\end{table}

\subsection{Physics beyond the SM} \label{sec:Limits}

A light gauge boson that couples to leptons and has a mass below approximately 70\GeV
may be produced on shell in $\PZ$ boson decays. If such a particle exists, it must modify $\BF$ with respect
to its SM predicted value.

Without focusing on specific models, we address this general possibility by hypothesizing a new light boson coupling significantly to leptons. This minimal extension of the SM was explored in Ref.~\cite{Rainbolt}, where the boson is labeled \PU. A scalar or vector \PU boson may mediate the $\ZtofL$ decay, as shown in Fig.~\ref{fig:NPFeynmanDiagrams}. The relevant parameters of \PU are its mass $m_\PU$ and the strengths of its couplings to leptons, $g_{\Pell}$. We consider cases where the \PU boson couples to electrons and muons equally~(\ie, $g_{\Pe} = g_{\PGm}$) and where it couples exclusively to leptons of one flavor~(\ie, $g_{\PGm} = 0$ or $g_{\Pe} = 0$).
In contrast to Ref.~\cite{EXO}, which targets a narrow peak in the dilepton invariant mass spectrum,
our procedure is insensitive to assumptions about the \PU boson total width $\Gamma_{\PU}$ because
we do not search for narrow structures in dilepton mass spectra.
Under the assumption that new physics appears as a narrow peak, the results
from Ref.~\cite{EXO} are expected to be more sensitive than the results presented here.

The 95\% confidence level (\CL) upper limit (UL) on individual channels $\BFfM$, $\BFtMtE$, $\BFfE$
and on the combined channel $\BF$ are calculated using the measured branching fractions given
in Table~\ref{tab:BSMMeasurements}.
The expected and observed upper limits are reported in Table~\ref{tab:Limits}.

In order to derive constraints on model parameters $m_{\PU}$ and $g_{\Pell}$, signal events are generated at LO
with {\MadGraph}~\cite{MadGraph} implementing the Lagrangian terms in Eqs.~\refeq{eqn:ScalarLagrangian}
and~\refeq{eqn:VectorLagrangian} using the {\FeynRules} package~\cite{FeynRules}.
The values for $m_{\PU}$ and $g_{\Pell}$ vary in the ranges $4 < m_\PU < 90\GeV$ and $0.005 < g_{\Pell} < 0.45$.
Each signal sample has $10^4$ events, including contributions from both the SM and the new physics process
and their interference. The \PU boson total width $\Gamma_{\PU}$ is calculated by \MADGRAPH.

For each sampled point in the $(m_{\PU}, g_{\Pell})$ parameter space, we calculate the acceptance $A^\NP$
of the fiducial region as defined in Section~\ref{sec:Events}. A sampled point is excluded when
\begin{equation} \label{eqn:Exclusion}
   A^\NP > \frac{\BF^\text{95\% CL UL}}{\BF} A^\SM,
\end{equation} 
where $\BF^\text{95\% \CL UL}$ is the 95\% \CL upper limit of $\BF$ listed in Table~\ref{tab:Limits}
and $A^\SM$ is the acceptance for SM events generated at LO with \MADGRAPH.

The results are presented as curves in the $(m_{\PU}, g_{\Pell})$ plane. These curves are obtained by
performing a fit in $(m_{\PU}, g_{\Pell})$ space and obtaining a smooth mapping
to $A^\NP$, that is, $f(m_{\PU}, g_{\Pell}) = A^\NP $.
The exclusion curve is given by the contour
\begin{equation*}
   f(m_{\PU}, g_{\Pell}) = \frac{\BF^\text{95\% CL UL}}{\BF} A^\SM,
\end{equation*} 
where the right side of the equality is the right side of the inequality in Eq.~\refeq{eqn:Exclusion}.

\begin{figure*} \centering
\includegraphics[width=\cmsFigWidth]{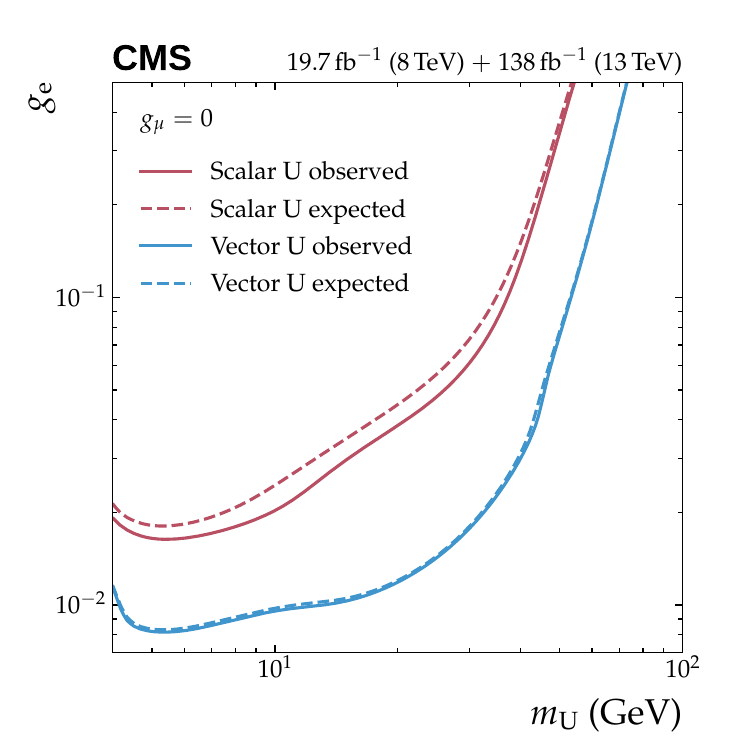} \hfill
\includegraphics[width=\cmsFigWidth]{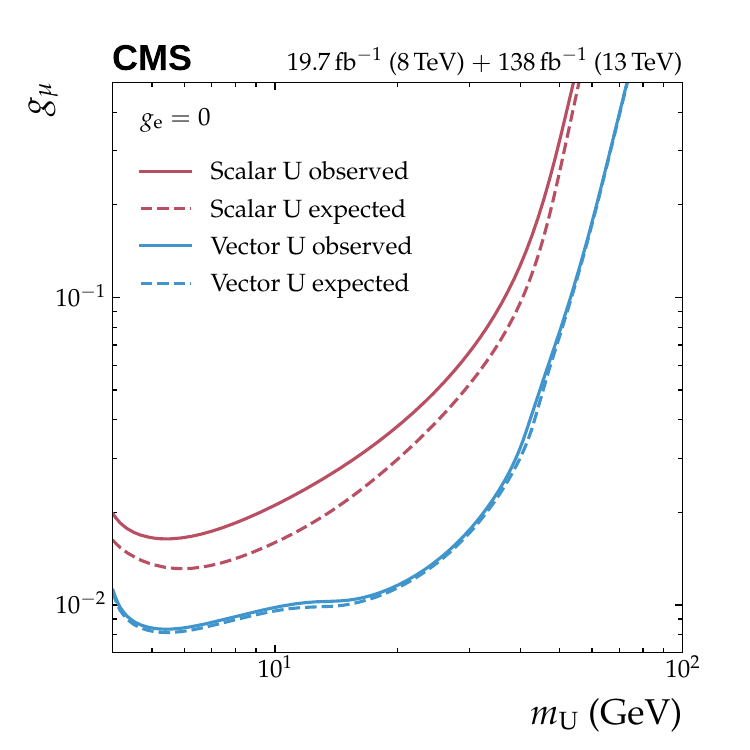}
\includegraphics[width=\cmsFigWidth]{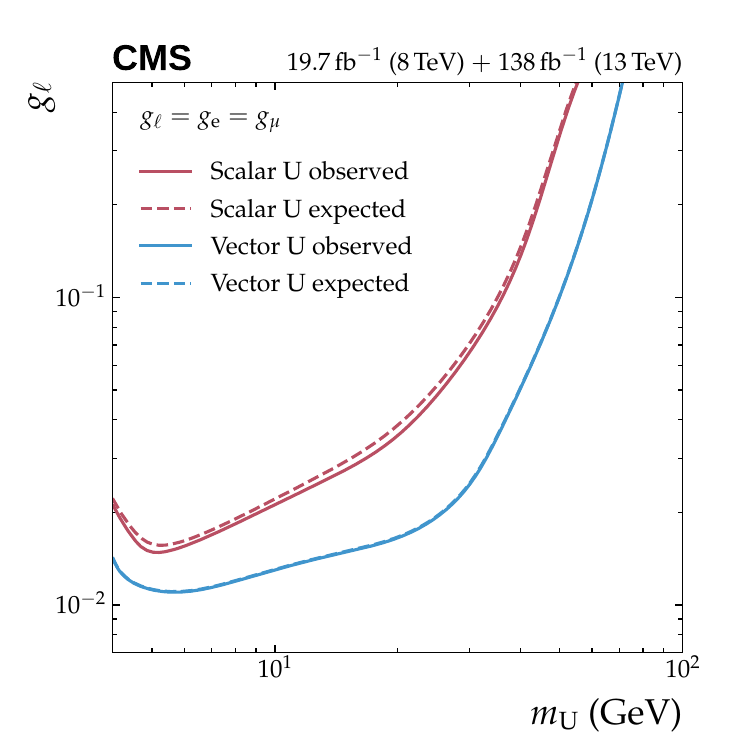} \hfill
\includegraphics[width=\cmsFigWidth]{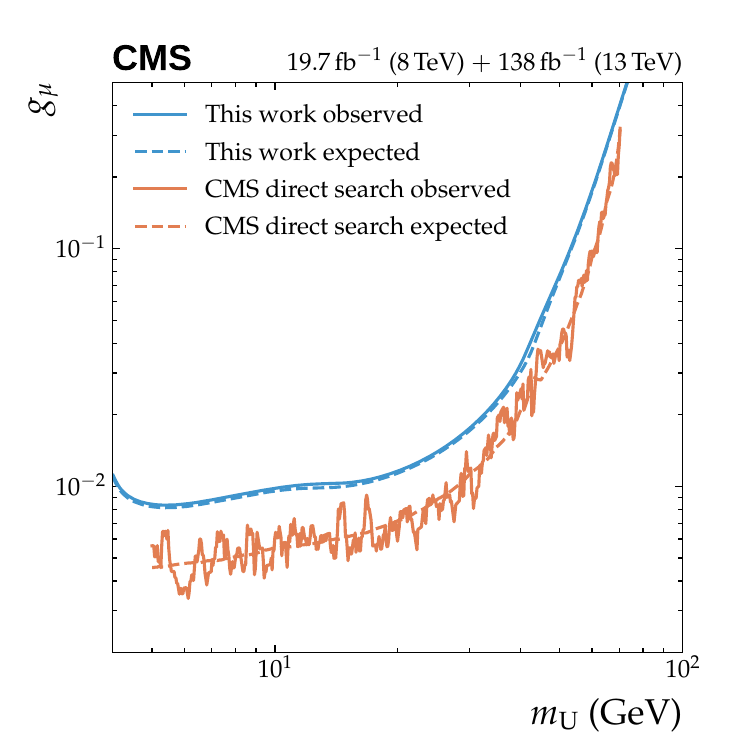}
\caption{Exclusion contours at 95\% CL for a BSM boson \PU:
(upper left) \PU couples only to electrons,
(upper right) \PU couples only to muons,
(lower left) \PU couples equally to electrons and muons.
The region above and to the left of the curve is excluded.
The lower right plot compares the exclusion contour obtain in this analysis to the
CMS direct search~\cite{EXO} for a vector $\PU$ boson that couples only to muons.
The dashed (solid) lines show the expected (observed) exclusion contours.}
\label{fig:Exclusion}
\end{figure*}

The excluded regions for the six models we consider are displayed in Fig.~\ref{fig:Exclusion}.
These constraints, based on the current analysis, are more stringent than those presented in Ref.~\cite{Rainbolt},
which employed a world average for $\BF$.
The strongest limits are found for the case of a vector $\PU$ with $g_{\Pe} = 0$, which complement the
CMS direct search for an ${L_\mu - L_\tau}$ vector gauge boson in the $\fM$ channel~\cite{EXO}.
The excluded regions from both analyses are shown together in Fig.~\ref{fig:Exclusion}~(lower right).
Similar results are obtained in Ref.~\cite{Drees2018}.

\begin{table} \centering
\topcaption{Upper limits at 95\%\,\CL on the branching fractions listed in Table~\ref{tab:BSMMeasurements}.}
\begin{scotch}{c c c}
	& \multicolumn{2}{c}{95\% CL upper limit [$\ten{$-6$}$]} \\
	Channel & Expected & Observed \\
\hline
	$\fM$ & 1.28 & 1.34 \\
	$\tMtE$ & 2.48 & 2.33 \\
	$\fE$ & 1.37 & 1.32 \\ [\cmsTabSkip]
	$\fL$ & 4.95 & 4.91 \\
\end{scotch}
\label{tab:Limits}
\end{table}

Other theoretical approaches for new physics based on effective field theories~\cite{Boughezal:EFT} can profit
from the constraints presented here, in particular, the differential decay distributions reported in
Section~\ref{sec:DifferentialDistributions}. The ATLAS Collaboration pursued this approach
in Ref.~\cite{ATLAS21}.

\section{Summary} \label{sec:Summary}
A detailed study of $\ZtofL$ decays (\Pell$ = $ \Pe or \PGm)
in proton-proton collisions at $\sqrt{s} = 8$ and 13\TeV has been presented.
The data correspond to an integrated luminosity of 19.7\fbinv collected at 8\TeV and
138\fbinv at 13\TeV with the CMS detector. A total of 1,877 events is observed in the $\ZtofL$ signal region.
The signal purity for each of the $\fM$, $\tMtE$, and $\fE$ final states is above~97\%.

A measurement of the branching fraction for the sum of all four-lepton decay modes has been reported:
$ \BF = ( 4.67 \pm 0.11 \stat \pm 0.10 \syst ) \times 10^{-6}$.
This measurement is more precise than all previous CMS and ATLAS results~\cite{CMS12, CMS16, CMS18, ATLAS14, ATLAS19, ATLAS21}
and has a combined statistical and systematic precision of 3.2\%. Measurements of the branching fractions for
the decays $\ZtofM$, $\ZtofE$, and $\ZtotMtE$ have also been reported. All measured branching fractions are consistent with
standard model (SM) expectations.

Measurements of differential decay rates for the $\ZtofL$ decay process as functions of nine kinematic and angular
quantities were presented. These measurements are reported in terms of the $\ZtofL$ partial width $\PWfL$. They are
unfolded to correct for the effects of detector resolution and scaled to the full phase space region defined by
the four-lepton invariant mass $80 < \mfL < 100\GeV$ and the dilepton invariant mass $\mLL > 4\GeV$.
These differential decay rates characterize the behavior of the $\ZtofL$
decay process and are consistent with simulations based on the SM.

In the interest of exploring possible contributions of new physics to the $\ZtofL$ channel,
a triple-product asymmetry measurement
was performed. The measurement was made possible by the high statistical precision of the analysis, and may be used
to probe for possible violations of charge conjugation and parity invariance in this decay.
This asymmetry is observed to be consistent with zero for all $\fL$ channels, as expected in the SM.
Finally, the $\BFfM$, $\BFfE$, and $\BFtMtE$ measurements were used to set limits on
new physics in the form of a scalar or vector boson that may mediate the $\ZtofL$ process. These limits are more stringent
than those set using previous measurements of $\BF$~\cite{Rainbolt} and complement direct searches for a narrow-width vector
boson decaying to a $\MM$ pair~\cite{EXO}.

\begin{acknowledgments}
We congratulate our colleagues in the CERN accelerator departments for the excellent performance of the LHC and thank the technical and administrative staffs at CERN and at other CMS institutes for their contributions to the success of the CMS effort. In addition, we gratefully acknowledge the computing centers and personnel of the Worldwide LHC Computing Grid and other centers for delivering so effectively the computing infrastructure essential to our analyses. Finally, we acknowledge the enduring support for the construction and operation of the LHC, the CMS detector, and the supporting computing infrastructure provided by the following funding agencies: SC (Armenia), BMBWF and FWF (Austria); FNRS and FWO (Belgium); CNPq, CAPES, FAPERJ, FAPERGS, and FAPESP (Brazil); MES and BNSF (Bulgaria); CERN; CAS, MoST, and NSFC (China); MINCIENCIAS (Colombia); MSES and CSF (Croatia); RIF (Cyprus); SENESCYT (Ecuador); ERC PRG and PSG, TARISTU24-TK10 and MoER TK202 (Estonia); Academy of Finland, MEC, and HIP (Finland); CEA and CNRS/IN2P3 (France); SRNSF (Georgia); BMFTR, DFG, and HGF (Germany); GSRI (Greece); MATE and NKFIH (Hungary); DAE and DST (India); IPM (Iran); SFI (Ireland); INFN (Italy); MSIT and NRF (Republic of Korea); MES (Latvia); LMTLT (Lithuania); MOE and UM (Malaysia); BUAP, CINVESTAV, CONACYT, LNS, SEP, and UASLP-FAI (Mexico); MOS (Montenegro); MBIE (New Zealand); PAEC (Pakistan); MSHE, NSC, and NAWA (Poland); FCT (Portugal); MESTD (Serbia); MICIU/AEI and PCTI (Spain); MOSTR (Sri Lanka); Swiss Funding Agencies (Switzerland); MST (Taipei); MHESI (Thailand); TUBITAK and TENMAK (T\"{u}rkiye); NASU (Ukraine); STFC (United Kingdom); DOE and NSF (USA).

\hyphenation{Rachada-pisek} Individuals have received support from the Marie-Curie programme and the European Research Council and Horizon 2020 Grant, contract Nos.\ 675440, 724704, 752730, 758316, 765710, 824093, 101115353, 101002207, 101001205, and COST Action CA16108 (European Union); the Leventis Foundation; the Alfred P.\ Sloan Foundation; the Alexander von Humboldt Foundation; the Science Committee, project no. 22rl-037 (Armenia); the Fonds pour la Formation \`a la Recherche dans l'Industrie et dans l'Agriculture (FRIA) and Fonds voor Wetenschappelijk Onderzoek contract No. 1228724N (Belgium); the Beijing Municipal Science \& Technology Commission, No. Z191100007219010, the Fundamental Research Funds for the Central Universities, the Ministry of Science and Technology of China under Grant No. 2023YFA1605804, the Natural Science Foundation of China under Grant No. 12535004, and USTC Research Funds of the Double First-Class Initiative No.\ YD2030002017 (China); the Ministry of Education, Youth and Sports (MEYS) of the Czech Republic; the Shota Rustaveli National Science Foundation (Georgia); the Deutsche Forschungsgemeinschaft (DFG), among others, under Germany's Excellence Strategy -- EXC 2121 ``Quantum Universe" -- 390833306, and under project number 400140256 - GRK2497; the Hellenic Foundation for Research and Innovation (HFRI), Project Number 2288 (Greece); the Hungarian Academy of Sciences, the New National Excellence Program - \'UNKP, the NKFIH research grants K 131991, K 138136, K 143460, K 143477, K 147557, K 146913, K 146914, K 147048, TKP2021-NKTA-64, and 2025-1.1.5-NEMZ\_KI-2025-00004, and MATE KKP and KKPCs Research Excellence and Flagship Research Groups grants (Hungary); the Council of Science and Industrial Research, India; ICSC -- National Research Center for High Performance Computing, Big Data and Quantum Computing, FAIR -- Future Artificial Intelligence Research, and CUP I53D23001070006 (Mission 4 Component 1), funded by the NextGenerationEU program, the Italian Ministry of University and Research (MUR) under Bando PRIN 2022 -- CUP I53C24002390006, PRIN PRIMULA 2022RBYK7T (Italy); the Latvian Council of Science; the Ministry of Science and Higher Education, project no. 2022/WK/14, and the National Science Center, contracts Opus 2021/41/B/ST2/01369, 2021/43/B/ST2/01552, 2023/49/B/ST2/03273, and the NAWA contract BPN/PPO/2021/1/00011 (Poland); the Funda\c{c}\~ao para a Ci\^encia e a Tecnologia (Portugal); the National Priorities Research Program by Qatar National Research Fund; MICIU/AEI/10.13039/501100011033, ERDF/EU, ``European Union NextGenerationEU/PRTR", projects PID2022-142604OB-C21, PID2022-139519OB-C21, PID2023-147706NB-I00, PID2023-148896NB-I00, PID2023-146983NB-I00, PID2023-147115NB-I00, PID2023-148418NB-C41, PID2023-148418NB-C42, PID2023-148418NB-C43, PID2023-148418NB-C44, PID2024-158190NB-C22, RYC2021-033305-I, RYC2024-048719-I, CNS2023-144781, CNS2024-154769 and Plan de Ciencia, Tecnolog{\'i}a e Innovaci{\'o}n de Asturias, Spain; the Chulalongkorn Academic into Its 2nd Century Project Advancement Project, the National Science, Research and Innovation Fund program IND\_FF\_68\_369\_2300\_097, and the Program Management Unit for Human Resources \& Institutional Development, Research and Innovation, grant B39G680009 (Thailand); the Eric \& Wendy Schmidt Fund for Strategic Innovation through the CERN Next Generation Triggers project under grant agreement number SIF-2023-004; the Kavli Foundation; the Nvidia Corporation; the SuperMicro Corporation; the Welch Foundation, contract C-1845; and the Weston Havens Foundation (USA).
\end{acknowledgments}\section*{Data availability} Release and preservation of data used by the CMS Collaboration as the basis for publications is guided by the  \href{https://doi.org/10.7483/OPENDATA.CMS.1BNU.8V1W}{CMS data preservation, re-use and open access policy}.

\bibliography{auto_generated}

\cleardoublepage \appendix\section{The CMS Collaboration \label{app:collab}}\begin{sloppypar}\hyphenpenalty=5000\widowpenalty=500\clubpenalty=5000\input{SMP-19-007-public-authorlist.tex}\end{sloppypar}
\end{document}

%% file: SMP-19-007-public-authorlist.tex
\cmsinstitute{Yerevan Physics Institute, Yerevan, Armenia}
{\tolerance=6000
A.~Tumasyan\cmsorcid{0009-0000-0684-6742}
\par}
\cmsinstitute{Institut f\"{u}r Hochenergiephysik, Vienna, Austria}
{\tolerance=6000
W.~Adam\cmsorcid{0000-0001-9099-4341}, J.W.~Andrejkovic, T.~Bergauer\cmsorcid{0000-0002-5786-0293}, S.~Chatterjee\cmsorcid{0000-0003-2660-0349}, K.~Damanakis\cmsorcid{0000-0001-5389-2872}, M.~Dragicevic\cmsorcid{0000-0003-1967-6783}, A.~Escalante~Del~Valle\cmsorcid{0000-0002-9702-6359}, R.~Fr\"{u}hwirth\cmsAuthorMark{1}\cmsorcid{0000-0002-0054-3369}, M.~Jeitler\cmsAuthorMark{1}\cmsorcid{0000-0002-5141-9560}, N.~Krammer\cmsorcid{0000-0002-0548-0985}, L.~Lechner\cmsorcid{0000-0002-3065-1141}, D.~Liko\cmsorcid{0000-0002-3380-473X}, I.~Mikulec\cmsorcid{0000-0003-0385-2746}, P.~Paulitsch, F.M.~Pitters, J.~Schieck\cmsAuthorMark{1}\cmsorcid{0000-0002-1058-8093}, R.~Sch\"{o}fbeck\cmsorcid{0000-0002-2332-8784}, D.~Schwarz\cmsorcid{0000-0002-3821-7331}, S.~Templ\cmsorcid{0000-0003-3137-5692}, W.~Waltenberger\cmsorcid{0000-0002-6215-7228}, C.-E.~Wulz\cmsAuthorMark{1}\cmsorcid{0000-0001-9226-5812}
% ...
\par}
\vskip\cmsinstskip
\dag:~Deceased\\
$^{1}$Also at TU Wien, Vienna, Austria\\
% ...
